\documentclass{aa}
\usepackage{graphicx}
\usepackage{txfonts}
\usepackage{gensymb}
\usepackage{wasysym}
\usepackage{caption}
\usepackage{subcaption}
\usepackage[normalem]{ulem}

\usepackage{natbib}
\bibpunct{(}{)}{;}{a}{}{,}

\usepackage[dvipsnames]{xcolor}
\usepackage{hyperref}
\hypersetup{
    colorlinks=true,
    linkcolor=MidnightBlue,
    filecolor=blue,      
    urlcolor=cyan,
    citecolor=TealBlue
}

\begin{document}

\title{RadioAstron reveals a change in the jet collimation profile of 3C\,84}

\author{P.~Benke \inst{1, 2}
        \and
        T.~Savolainen \inst{3, 4, 1}
        \and
        G.~Giovannini \inst{5, 6}
        \and
        Y.~Y.~Kovalev\inst{1}
        \and
        G.~Bruni\inst{7}
        \and
        M.~M.~Lisakov\inst{8}
        \and
        M.~Giroletti\inst{6}
        \and
        E.~Ros\inst{1}
}

\institute{Max Planck Institute for Radio Astronomy,
           Auf dem Hügel 69, D--53121 Bonn, Germany
           \and
           Julius Maximilians University W\"urzburg, Faculty of Physics and Astronomy, Institute for Theoretical Physics and Astrophysics, Chair of Astronomy, Emil-Fischer-Str. 31, D-97074 Würzburg, Germany
           \and
           Aalto University Department of Electronics and Nanoengineering,
           PL 15500, FI--00076 Aalto, Finland
           \and
           Aalto University Metsähovi Radio Observatory,
           Metsähovintie 114, FI--02540 Kylmälä, Finland
           \and
           Dipartimento di Fisica e Astronomia, Università di Bologna, via Gobetti 93/2, I--40129 Bologna, Italy
           \and
           INAF--IRA, Via Gobetti 101, I-40129 Bologna, Italy
           \and
           INAF -- Institute for Space Astrophysics and Planetology, via del Fosso del Cavaliere, 100, I-00133 Rome, Italy
           \and
           Instituto de F\'{i}sica, Pontificia Universidad Cat\'{o}lica de Valpara\'{i}so, Casilla 4059, Valpara\'{i}so, Chile
        }

\date{Received / Accepted }

\abstract{Due to its brightness and proximity, the radio galaxy 3C\,84 (optical counterpart NGC\,1275 in the Perseus cluster) has been the target of extensive studies investigating the central parsec region of its active galactic nucleus. In 2003, its most recent active phase resulted in a plasma ejection visible in the southern jet, which presented a unique opportunity to study jet formation and evolution at high angular resolution with very long baseline interferometry (VLBI).}
{We aim to study the morphology, evolution, and spectral properties of the restarted jet three years after the first ultra-high angular resolution observations with the RadioAstron space-VLBI satellite in September 2013.}
{To study 3C\,84, we used space-VLBI observations carried out in September 2016 at 22\,GHz with a global VLBI network and the 10 m Spektr-R radio telescope in orbit as well as quasi-simultaneous multifrequency observations at 4.8, 8, 15, and 43\,GHz from the Very Long Baseline Array, including the Effelsberg 100 m telescope.}
{We present the 22\,GHz RadioAstron image of 3C\,84 from 2016, which reveals the source's central region at a 58~$\mu$as effective resolution. During the three years that elapsed between the first and second space-VLBI observations, the source underwent significant morphological changes. We confirm the existence of the limb-brightened jet and counter-jet reported earlier as well as a flip in the position of the hotspot discovered recently via VLBI monitoring at 43\,GHz. Based on measuring the collimation profile, we find that it has evolved from being quasi-cylindrical to parabolic. This is most likely the result of the decreased pressure of the mini-cocoon, which was inflated by the jet and contains hot gas that cannot confine the jet efficiently as it propagates further away from the core. Finally, we also constrained the magnetic field strength in the core region and the hotspot.}
{}


\maketitle

\section{Introduction}

Although the first active galaxy was discovered more than a hundred years ago \citep{1918PLicO..13....9C}, several aspects of these powerful and luminous sources are still not well understood. Radio-loud active galactic nuclei (AGNs) exhibit collimated outflows of relativistic plasma called jets, whose launching, acceleration, and collimation mechanisms still pose open questions for us to investigate \citep{1997ARA&A..35..607Z,2019ARA&A..57..467B}. Due to the synchrotron radiation of relativistic electrons, jets can be studied at very high angular resolution in the radio domain using very long baseline interferometry (VLBI) at either short wavelengths \citep{2017A&ARv..25....4B} or with extremely long baselines. Observations of the central region of M\,87 with the Event Horizon Telescope at $1.3$~mm \citep{eht19a} were capable of reaching an angular resolution of 20~$\mu$as and detecting the shadow of the central black hole. The angular resolution using space VLBI observations with the RadioAstron telescope could exceed even this extraordinary resolution and resolve water masers in NGC\,4258 at a resolution of  11~$\mu$as \citep{2022NatAs...6..976B} and detect the blazar OJ\,287 down to 12~$\mu$as fringe spacing \citep{2022ApJ...924..122G}.

The low-power radio galaxy 3C\,84 (NGC\,$1275$, Perseus\,A) is located at the center of the Perseus cluster at $z=0.0176$, 75.7~Mpc from us \citep{strauss92}. It is not only a favorable target because of its proximity, but also because early VLBI observations have already revealed the variable nature of this source: \citet{pauliny66} reported an increase in flux density and angular diameter in the 1960s. Notably, 3C\,84 has exhibited several periods of activity during the past decades \citep[e.g.,][]{nesterov95}. Its latest activity started in 2003, when a new component emerged from the radio core \citep{suzuki12}. This years-long activity coincided with the first detection of $\gamma$-ray emission from 3C\,84 by the \textit{Fermi} Gamma-ray Space Telescope \citep{abdo09} in 2008. The source was detected at very high energies ($E > 100$~GeV) by the MAGIC telescopes in August 2010 \citep{aleksic12}, and a bright flare was observed on the night of December 31, 2016 \citep{magic18}. While the $\gamma$-ray emission of 3C\,84 is variable, it is regularly detected by the \textit{Fermi} LAT. However, very high energy detections are limited only to the flaring episodes exhibited by the source. While no significant correlation has been found between the radio and $\gamma$-ray activity and the location of the emission regions in these two energy regimes, a common trend is present \citep{2021ApJ...914...43H}. Nevertheless, this recently emerged bright component provides an ideal case study to investigate the launching mechanism and evolution of parsec-scale jets at extremely high resolution with RadioAstron.

RadioAstron was a space-based VLBI project led by the Astro Space Center of the Lebedev Physical Institute. The 10 m space radio telescope (SRT) on board the Spektr-R spacecraft designed by the Lavochkin Association was launched in $2011$. The telescope was decommissioned after 7 years of operation on May 30, 2019. The orbit of RadioAstron could reach an apogee height of $370,000$~km. The telescope operated in the K ($1.19 - 1.63$~cm), C (6.2~cm), L (18~cm), and P bands (92~cm), and it could reach the nominal resolution of 7~$\mu$as at the highest frequency and longest baselines \citep{kardashev13}. This extraordinary resolution enabled the investigation of AGN jets close to their launching region in nearby AGNs, including 3C\,84.

The first observations of 3C\,84 with RadioAstron at 22\,GHz and 5\,GHz were carried out in September 2013, and the results were published in \citet{giovannini18} and \citet{savolainen21}, respectively. \citet{giovannini18} reported the existence of a limb-brightened jet that had also been observed by \citet{nagai14} and a limb-brightened counter-jet. The collimation profile ($r\propto z^{a}$, where $r$ is the jet width, $z$ is the de-projected distance from the core, and $a$ is a power-law index) measurement revealed a quasi-cylindrical jet with a power-law index of $0.17\pm0.01$. Another important finding of this observation was the measured wide jet base, as the restarted jet is $r\approx250 r_{\mathrm{g}}$ wide at only $z\approx350r_{\mathrm{g}}$ from the radio core. This suggests either that the jet or its sheath was launched from the accretion disk or that the jet expanded laterally rather quickly in the first few hundred gravitational radii from the core.

Our goal is to study the morphology, evolution, and spectral properties of the restarted jet of 3C\,84 in 2016, three years after the first 22\,GHz RadioAstron observations in 2013. We present imaging results from the 2016 RadioAstron observations as well as a collimation profile measurement. Using the quasi-simultaneous multifrequency Very Long Baseline Array (VLBA) data recorded together with the RadioAstron observations, we performed spectral analysis of the core and hotspot regions and measured the core shift. This enabled us to probe the magnetic field strength in the restarted jet and to measure the distance to the jet apex, as differing values have been reported in the literature. In this paper, we first present the observations and the methods used for data reduction in Sect.~\ref{obs}. Then we proceed with discussion of our analysis and results on the evolution of the restarted jet in Sect.~\ref{jet_params} and Sect.~\ref{morphology}. We present our collimation profile measurement, spectral analysis, and core shift estimation in Sect.~\ref{collimation}, Sect.~\ref{spectral}, and Sect.~\ref{coreshift}. Finally, we summarize our work in Sect.~\ref{conclusion}.

In this work we assume a $\Lambda$ cold dark matter cosmology with $H_0=70.7$~km~s$^{-1}$~Mpc$^{-1}$, $\Omega_{\Lambda}=0.73$, and $\Omega_{\mathrm{M}}=0.27$. At $z=0.0176$, this corresponds to a scale of 0.354~pc/mas. While there is a wide range of black hole masses and jet inclination angle estimates for 3C\,84 \citep[see for example][]{scharwachter13, 2009AJ....138.1874L, 2017MNRAS.465L..94F}, to be compatible with \citet{giovannini18}, we used a black hole mass of $2\times10^9$~M$_{\astrosun}$ and inclination angle of $18$\degree. (See \citet{giovannini18} for the discussion of this choice.) This means that $1~\mathrm{mas}\approx3.58\times10^3~\mathrm{r_g}$.

\section{Observations and data reduction}
\label{obs}

\subsection{Calibration and imaging of RadioAstron space VLBI data}

Space-VLBI observations at 22\,GHz were carried out using a global VLBI network of 23 ground-based antennas and the SRT, RadioAstron. The 23 ground-based antennas include the 10 VLBA stations (Brewster, Fort Davis, Hancock, Kitt Peak, Los Alamos, Mauna Kea, North Liberty, Owens Valley, Pie Town, and St. Croix) and the 3 Korean VLBI Network stations (Tamna, Ulsan, and Yonsei), as well as Badary, the Effelsberg 100 m telescope, Hartebeesthoek, Medicina, Mets\"ahovi, Noto, Onsala, Svetloe, Yebes, and Zelenchukskaya. Apart from the 24 antennas that provided data, the observation was also scheduled at Shanghai, Toru\'n, Urumqi, and Robledo, but no usable data were gathered at those stations. The observations started on September 11, 2016, at 13~UT, and finished after 30~hours on September 12, 2016. In addition to the target, 0300+470 and 1823+568 were observed, the first source as a D-term and amplitude calibrator, and the second one as an electric vector position angle calibrator. The data were recorded with dual circular polarization at two sub-bands (intermediate frequencies) with 16~MHz bandwidth. Similarly to the 2013 observations, the recording rate of RadioAstron was 128~Mbps, while the ground array antennas recorded data at a 256~Mbps rate. RadioAstron used a 1~bit sampling, while the ground array telescopes used 2~bit sampling, resulting in a difference in the bit rate between the ground and space antennas. The correlation was performed at the Max-Planck-Institut f\"ur Radioastronomie in Bonn, Germany, with a DiFX software correlator modified to allow for correct calculation of the path delays for an orbiting antenna while taking into account special and general relativistic effects \citep{deller07, deller11, 2016Galax...4...55B}. The correlator integration time was 0.25~s. 

Since the SRT is a 10 m antenna with passively cooled receivers, its sensitivity is not always sufficient to detect the source on single baselines between a ground-based antenna and RadioAstron. In principle, the detection threshold in fringe search on ground-space baselines can be lowered by phasing up the antennas in the ground array. However, according to \citet{kogan96}, the same increase in the fringe search sensitivity on the ground-space baselines can be obtained by employing global fringe fitting \citep{schwab83} if the ground-ground baselines are much more sensitive than the ground-space baselines. In global fringe fitting, we use all the available data and determine delay, rate, and phase residuals per antenna, not per baseline.  

\begin{figure*}[h!]
    \centering
    \includegraphics[width=0.8\linewidth]{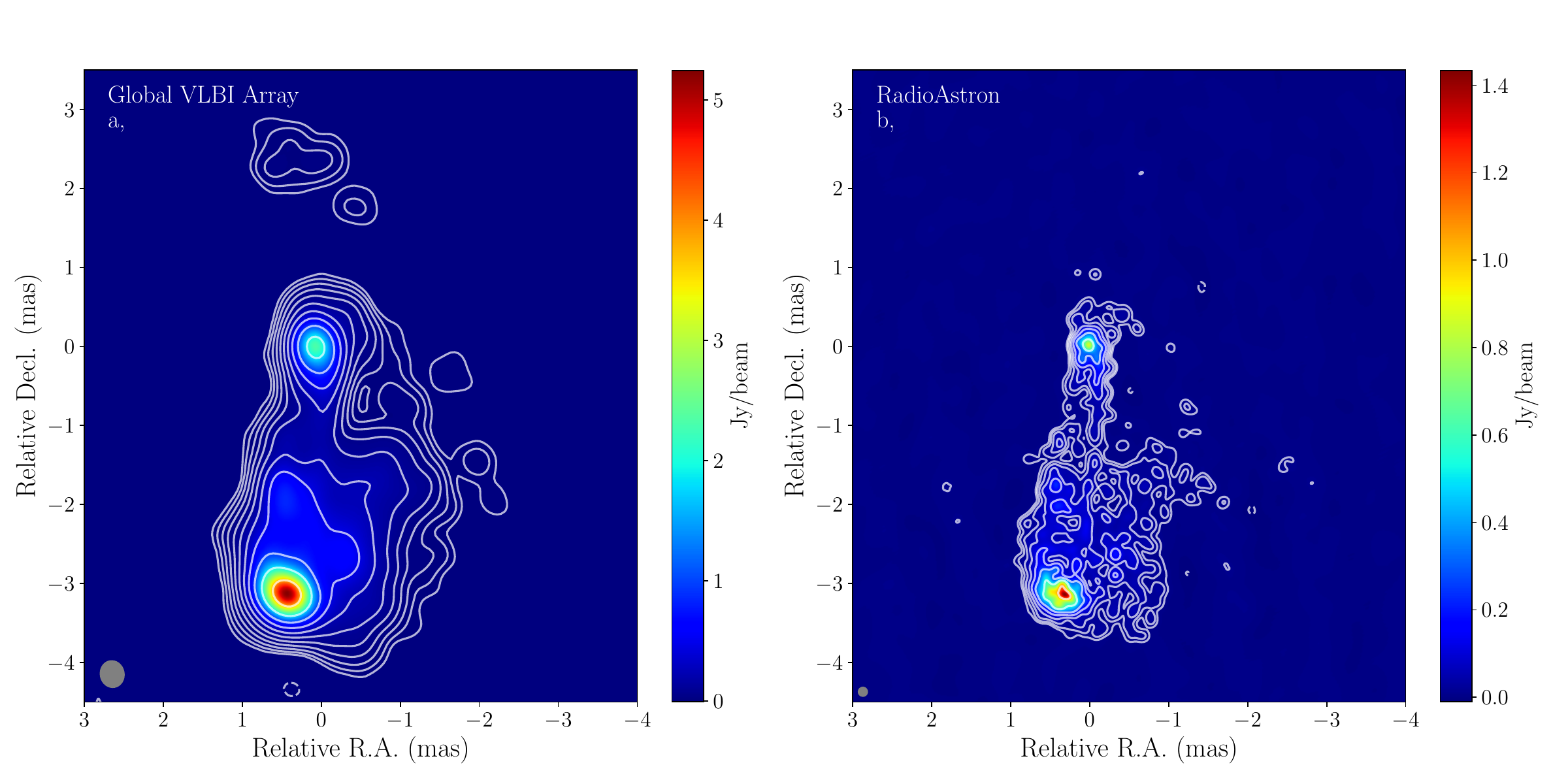}
    \caption{Hybrid images of 3C\,84 from the $2016$ RadioAstron data. The core has been shifted to the map center in both images based on the values listed in Table~\ref{tab:cs_comps}. Panel~\textit{a} shows the image obtained from the ground array data, whereas panel~\textit{b} contains data from the SRT as well. The lowest contours are at 3.7~mJy/beam for panel~\textit{a} and at 8.6~mJy/beam for panel~\textit{b,} and further contours increase by a factor of two. The image of the global array data has a total flux density of $27.8\pm1.4$~Jy, and the half-power beam width of the restoring beam is $0.35\times0.32$~mas at PA of $10.8\degr$. The rms noise level is 0.7~mJy/beam. The super-resolved RadioAstron image has a half-power beam width of $0.13\times0.13$~mas. The rms noise level is 1.9~mJy/beam, and the total flux density is $23.0\pm2.3$~Jy.}
    \label{fig:cln}
\end{figure*}

The calibration was carried out in the National Radio Astronomy Observatory's Astronomical Image Processing System (\texttt{AIPS}), and imaging of the data was performed in \texttt{Difmap} \citep{shepherd94}. Data reduction was first performed for the ground array only, and the resulting source model (see the left panel of Fig.~\ref{fig:cln}) was used to correct for the resolved source structure while fringe fitting the ground array data for the second time. After this, the ground-array data are self-calibrated, and we can proceed to find a signal on the ground-space baselines. Since we have removed most of the atmospheric phase fluctuations in self-calibration, the solution intervals could be increased significantly and further enhance the sensitivity on the ground-space baselines (for more discussion about the coherence times on the ground-space baselines, see \citealt{savolainen21}). In the \texttt{AIPS} task \texttt{FRING}, we used a weighted average of up to three baseline combinations in the initial coarse fringe search stage to improve the sensitivity. While the signal-to-noise  ratio (S/N) cutoff in the initial fringe search is generally set to 5, the low S/N space VLBI data often requires examining lower detection thresholds \citep{savolainen21}. This makes false detections more likely (see Appendix~\ref{fdr}) unless one limits the size of the search window in the delay and rate space. Limiting the search window size requires that at least some fringe detections on the space-ground

\begin{figure}[h!]
\centering
\includegraphics[width=0.8\linewidth]{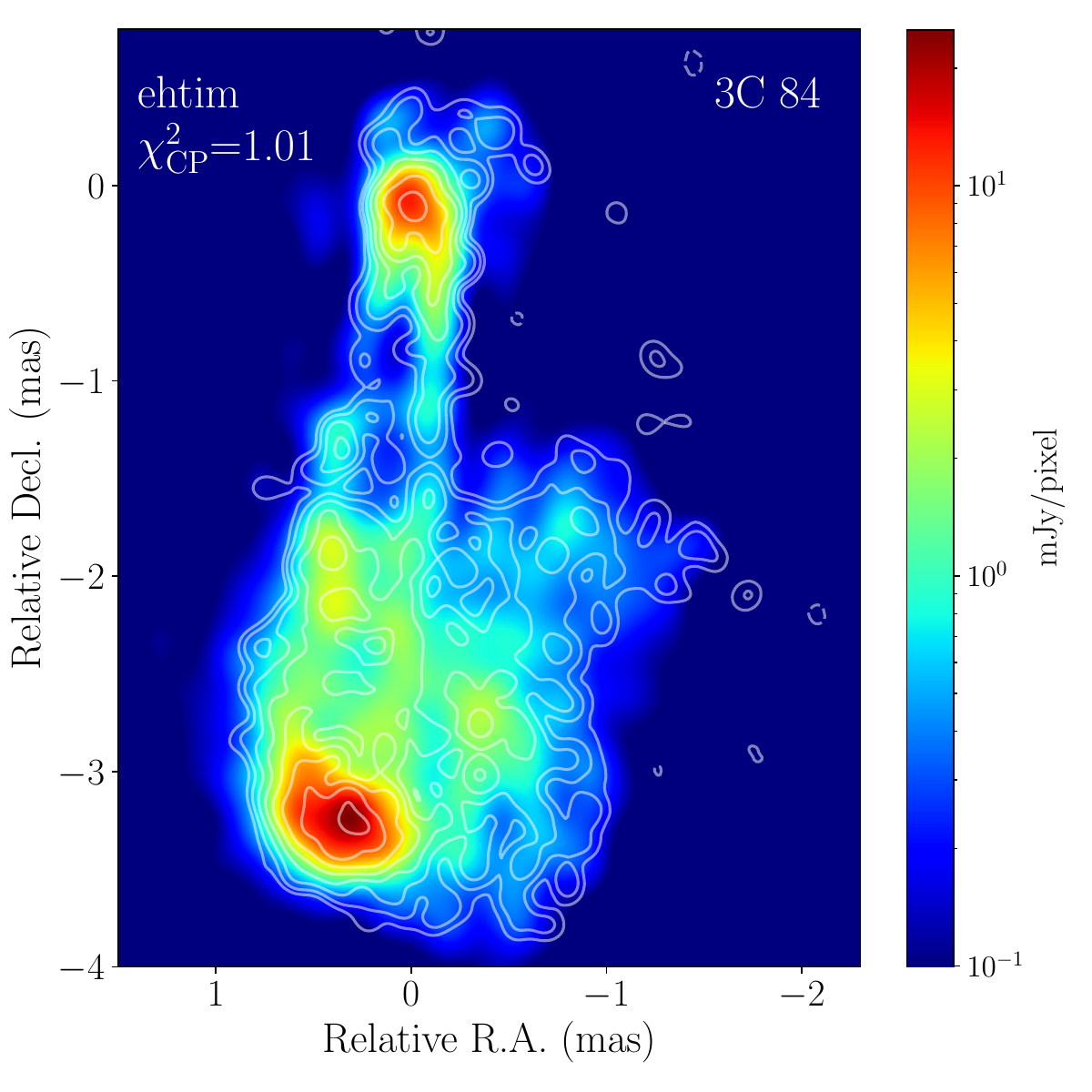}
    \caption{Final image obtained via \texttt{ehtim} from the $22$\,GHz RadioAstron dataset overlaid with the contours of the \texttt{clean} image for comparison. The reduced $\chi^2$ of the closure phases is indicated in the upper-left corner. The core was shifted to the map center based on the values shown in Table~\ref{tab:cs_comps}.}
    \label{fig:ehtim}
\end{figure}

\begin{figure}
\includegraphics[width=0.8\linewidth]{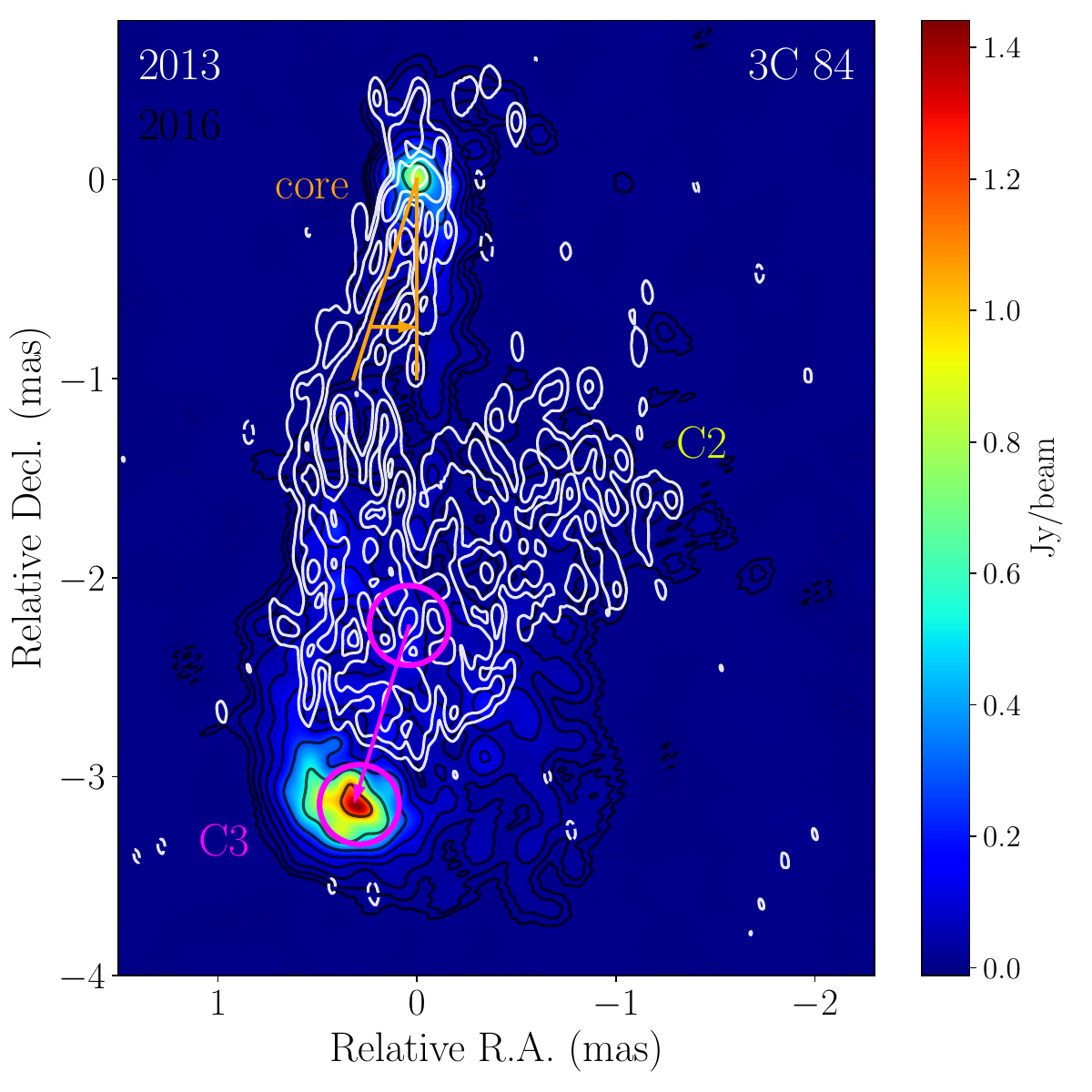}
    \caption{RadioAstron images of 3C\,84 from $2013$ (white contours) and $2016$ (color map and black contours). The core was shifted to the map center based on the values shown in Table~\ref{tab:cs_comps}. Black contours are the same as in Fig.~\ref{fig:cln} panel \textit{b}, and white contours start from $0.75$~Jy/beam and increase by a factor of four. Magenta circles mark the change in the hotspot (C3) position between the two epochs, and orange lines indicate the $32\pm1$\degree change in the jet position angle.}
    \label{fig:over}
\end{figure}

\noindent
baselines exist, and the fringes have been shifted close to zero rate and delay. Frequency averaging in sub-bands after removing instrumental delays and combining left and right circular polarizations can also increase the chances of finding the signal. 

During the fringe search at the correlator, the \texttt{PIMA} software \citep{petrov11} found fringes between the Effelsberg $100$~m telescope and RadioAstron in four scans, between 23:15 and 00:15~UT with S/Ns in the range of 7.6 to 8.4. During these fringe detections, the SRT was far away from the perigee of its orbit, so acceleration was not fit. We could recover the signal in \texttt{FRING} in all four scans, but we had no fringe detections in the rest of the scans, including RadioAstron. However, two of the four fringe detections had a rather low S/N. These two detections were justified by the smoothly changing rates and delays, and the \texttt{PIMA} detection.

Imaging RadioAstron data requires a careful treatment. To fully utilize the high angular resolution provided by RadioAstron, the ground array is down-weighted by using super-uniform weighting (\texttt{uvweight 5,-1}) in \texttt{Difmap}. In order not to create spurious flux from the noisy data of the SRT \citep{savolainen21,martividal08}, the solution interval for phase self-calibration for RadioAstron was kept at $1$~min. Since RadioAstron is not affected by the atmosphere, only a single gain correction factor was determined for the SRT, and further steps of amplitude self-calibration with decreasing solution intervals were only performed for the ground array. This single gain correction is also supported by the excellent pointing stability of the SRT, which has a measured standard deviation of just 0.24\arcsec \citep[Sect. 5.6.3 in][]{2014CosRe..52..365L} for the beam size of $6\arcmin\times13\arcmin$ at 22\,GHz \citep[Table~2 of][]{kardashev13}. To test the robustness of the counter-jet, we mapped the data again in the same manner, but without placing \texttt{clean} windows at the location of the feature. We found that we could not self-calibrate the counter-jet away, and it consistently appeared on the residual map through the imaging with a brightness exceeding 0.03 Jy/beam. It is also pertinent to note that the $\chi^2$ of the closure phase is 2 in the case when the counter-jet is cleaned, and 4 if it is not, indicating that it is not an artifact. The final hybrid images are displayed in Fig.~\ref{fig:cln}.

\begin{figure}[h!]
    \centering
    \includegraphics[width=\linewidth]{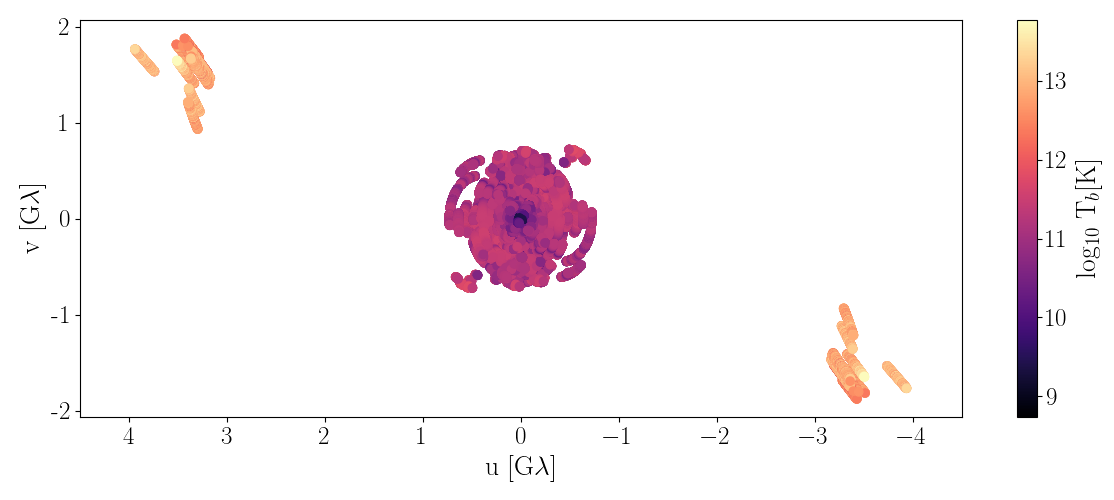}
    \caption{($u$,$v$) coverage and minimum brightness temperatures estimated from the interferometric visibilities as described in Sect.~\ref{jet_params} and Eq.~$2$ \citep{lobanov15}.}
    \label{fig:tb}
\end{figure}

Since the $(u,v)$ coverage of space baselines is very limited, making it easy to create spurious features in the imaging process, we tested the significance of the image features by imaging the data also with the \texttt{ehtim} package \citep{chael16} using a regularized maximum likelihood method. The image was solved pixel by pixel, and the final image, $\textbf{I}$, was obtained by minimizing the objective function
\begin{equation}
J(\textbf{I}) = \Sigma_{\mathrm{data terms}} \alpha_{\mathrm{D}} \chi_{\mathrm{D}}^2 (\textbf{I}, \textbf{d}) - \Sigma_{\mathrm{regularizers}} \beta_{\mathrm{R}} S_{\mathrm{R}} (\textbf{I}),
\end{equation}
where $\chi_{\mathrm{D}}^2$ are the goodness-of-fit terms for the data vector $\textbf{d}$ and image $\textbf{I}$, $S_{\mathrm{R}} (\textbf{I})$ are the regularize terms, and the weights of data and regularizer terms are specified by $\alpha_{\mathrm{D}}$ and $\beta_{\mathrm{R}}$. As a prior image, we used the ground array \texttt{CLEAN} image restored with a $0.5$~mas circular beam. We performed the imaging similarly as described in \citet{savolainen21}, first including data terms of closure phases, log closure amplitudes and visibility amplitudes with low weights. After this, we self-calibrated the phases with a solution interval of $10$~s for the ground-based telescopes, and $1$~min for RadioAstron. In the second round of imaging, we used both closure quantities and complex visibilities, then self-calibrated visibility amplitudes with a solution interval of $30$~s for the ground array telescopes, and in the case of RadioAstron for the whole length of the observation. The final imaging round was performed using the same data terms as before but with updated weights. We used the maximum entropy, total variation, and total squared variation regularizer terms, as well as a regularizer of the total flux density with the respective weights of 100, 1, 1, and $10^4$. We performed no parameter survey, as we found that variations in the regularizer weights did not significantly affect the final image nor the $\chi^2$ values of the data terms. The final image is shown in Fig.~\ref{fig:ehtim}.

When comparing the \texttt{CLEAN} map in Fig.~\ref{fig:cln} with the maximum likelihood image made with \texttt{ehtim} in Fig.~\ref{fig:ehtim}, we noted that the latter produces smoother features than \texttt{CLEAN} does. As shown recently in \citet{2023NatAs...7.1359F}, \texttt{ehtim} is better at recovering filamentary structures, such as limb-brightening in the jet and counter-jet. Apart from general differences arising from the distinct nature of these two algorithms, the source shows similar structures (and fits the observed data) on both maps, supporting the robustness of the existence of the edge-brightened features.

\subsection{Reduction of the ground-based multifrequency data}

Simultaneous multifrequency data at 4.8, 8, 15, and 43\,GHz were observed with the VLBA and the Effelsberg 100 m Telescope on September 12, 2016, between 05:00 and 18:30~UT. The Medicina 32~m telescope also joined the $4.8$~GHz observations. The data were correlated with the DiFX software correlator \citep{deller07, deller11} at the Max-Planck-Institut für Radioastronomie, with dual circular polarization at sub-bands of $16$~MHz bandwidth each and a recording rate of 256~Mbps. The data reduction was performed in \texttt{AIPS}, where we applied a priori amplitude calibration based on the system noise temperature, $T_{\mathrm{sys}}$ and elevation-dependent gain curves provided by the stations. This step also applied opacity corrections based on the system temperature measurements, fit receiver temperatures, and weather data recorded at each antenna. Residual delay and rate fitting were performed with the task \texttt{FRING}. Bandpass corrections were not applied. Effelsberg data were lost at 43\,GHz due to technical issues. Imaging was performed in a hybrid manner in \texttt{Difmap} \citep{shepherd94}, iterating the deconvolution method \texttt{clean} and self-calibration, shortening the solution interval for each amplitude calibration step. The resulting hybrid images are displayed in Fig.~\ref{fig:ground_cln}.

 \begin{table*}[h]
 \centering
       \caption[]{Brightness temperature, Lorentz, and Doppler factor measurements for the core and hotspot.}
       \fontsize{9pt}{9pt}\selectfont
         \begin{tabular}{lccccc}
            \hline\hline
            Component      &  $\mathrm{T}_{\mathrm{b, obs}}$~[K]\tablefootmark{a} & $\Gamma_{\mathrm{T_b}}$\tablefootmark{b} & $\delta_{\mathrm{T_b}}$\tablefootmark{c} & $\Gamma_{\mathrm{kin}}$\tablefootmark{d} & $\delta_{\mathrm{kin}}$\tablefootmark{e} \\
            \hline
            Core & $(1.9\pm3.1)\times10^{12}$ & $4.9\pm0.5$ & $9.7\pm1.0$ & - & -\\
            Hotspot & $(6.7\pm0.7)\times10^{11}$ & $1.8\pm0.2$ & $3.4\pm0.3$ & $1.2\pm0.1$ & $1.8\pm0.1$ \\
            \hline
         \end{tabular}
         \tablefoot{
         The viewing angle was assumed to be 18\degr.
         \tablefoottext{a}{Brightness temperature.}
         \tablefoottext{b}{Lorentz factor determined from the brightness temperature.}
         \tablefoottext{c}{Doppler factor determined from the brightness temperature.}
         \tablefoottext{d}{Lorentz factor determined from kinematics of the jet.}
         \tablefoottext{e}{Doppler factor determined from kinematics of the jet.}
         }
         \label{tab:Tb}
   \end{table*}

\section{Jet parameters}
\label{jet_params}

The main components of 3C\,84, marked in Fig.~\ref{fig:over}, are the core, which is connected to the hotspot (denoted as C3) by a limb-brightened jet. The faint, diffuse (denoted as C2 in the figure) to the southwest may be a remnant of previous activity in the source.

Brightness temperatures of the core and hotspot in the RadioAstron data were measured via two methods. First, the self-calibrated dataset was fit with circular and elliptical Gaussian components using the \texttt{Difmap} command \texttt{modelfit}. However, because of the sparse ($u$, $v$) coverage on space baselines and the complex structure of 3C\,84, we also use the method described in \citet{lobanov15}, where the minimum and maximum brightness temperature is estimated from the interferometric visibilities and their errors alone, under the assumption that the brightness distribution is circular or axially symmetric and for the maximum brightness temperature we also suppose that the observed structure is marginally resolved. We measured the minimum brightness temperature, $T_{\mathrm{{b, min}}}$ as \citep{lobanov15}
\begin{equation}
    T_{\mathrm{{b, min}}} [\mathrm{K}] = 3.09 \Bigg( \frac{b}{\mathrm{km}} \Bigg)^2 \Bigg( \frac{V_{\mathrm{q}}}{\mathrm{mJy}} \Bigg),
\end{equation}
where $b$ is the baseline length, and $V_{\mathrm{q}}$ is the visibility amplitude. Brightness temperature measurements from this method are shown in Fig \ref{fig:tb}. The average $T_{\mathrm{{b, min}}}$ on the space baselines is $6.4\times10^{12}$~K.

From the fit Gaussian components, the brightness temperature ($T_{\mathrm{b, obs}}$, see Table~\ref{tab:Tb}) was obtained as
\begin{equation}
    T_{\mathrm{b, obs}} [\mathrm{K}] = 1.22\times 10^{12} \Bigg(\frac{S_{\nu}}{\mathrm{Jy}}\Bigg) \Bigg(\frac{\nu}{\mathrm{GHz}}\Bigg)^{-2} \Bigg(\frac{b_{\mathrm{min}}\times b_{\mathrm{maj}}}{\mathrm{mas}^2}\Bigg)^{-1} (1+z),
\end{equation}
where $S_{\nu}$ is the flux density of the components, $\nu$ is the frequency of the observation and $b_{\mathrm{min}}$ and $b_{\mathrm{maj}}$ are the full width at half maximum (FWHM) sizes of the minor and major axis of the component. All component sizes are above the resolution limit \citep{kovalev05}. The intrinsic brightness temperature, $T_{\mathrm{b,int}}$, of a radio-emitting component is calculated as
\begin{equation}
    T_{\mathrm{b,int}} = \frac{T_{\mathrm{b,obs}}}{\delta(1+z)},
\end{equation}
where $\delta=(1-\beta^2)^{-1/2}/(1-\beta\cos{\theta})$ is the Doppler factor, $\beta$ is the component speed in the units of $c$, and $\theta$ is the jet viewing angle. Hence, by assuming an intrinsic brightness temperature value, we can obtain an estimate of the Doppler factor. Instead of the equipartition brightness temperature, $T_{\mathrm{eq}}\approx5\times10^{10}$~K \citep{readhead94}, we adopted $T_{\mathrm{b, int}}=2\times 10^{11}$\,K suggested by \citet{cohen07} for sources in their highest brightness states, which applies to 3C\,84 after the ejection of the restarted jet. This value, however, should be taken as a lower limit since AGNs during high states often exhibit intrinsic brightness temperatures greater than $2\times 10^{11}$\,K \citep{2006ApJ...642L.115H}. We inferred the Doppler factors for the radio core and the hotspot and the bulk Lorentz factor, $\Gamma=(1-\beta^2)^{-1/2}$, for the core and the hotspot.

Kinematic analysis has been performed by several groups since the emergence of the new jet in 2003. In the first few years after its appearance, the hotspot near the leading edge of the jet has accelerated to $\beta_{\mathrm{app}}\approx 0.2~c$ \citep{nagai10, suzuki12}, and \citet{2024ApJ...970..176K} has reported that C3 has on average, maintained this speed for about 20~yrs, with occasional acceleration or deceleration of its advancing. The 15\,GHz monitoring of the Monitoring Of Jets in Active galactic nuclei with VLBA Experiments (MOJAVE) team reports a maximum jet speed of $0.409\pm0.046~c$ and significantly accelerating components in the restarted jet of 3C\,84 \citep{2021ApJ...923...30L}. The apparent advance speed, $\beta_{\mathrm{app}}$ of the restarted jet was measured between the two RadioAstron observations by fitting elliptical Gaussian components in the ($u$, $v$) plane of the 2013 and 2016 observations with the \texttt{Difmap modelfit} task. The apparent speed is $\beta_{\mathrm{app}}=0.37\pm0.05$~$c$, corresponding to the advance speed of $\beta=0.56 \pm0.03$ assuming $\theta=18\degr$. However, viewing angle estimates up to $65\degr$ \citep{2017MNRAS.465L..94F} exist of the 3C\,84 jet, so $\beta$ ranges from $0.56\pm0.03$ to $0.35\pm0.04$. This speed, however, does not reflect the jet speed, as the hotspot moves slower than the flow due to its interaction with the ambient medium \citep{2021ApJ...920L..24K}. The bulk Lorentz factor, $\Gamma_{\mathrm{kin}}=1.2\pm0.1$, and the Doppler boosting factor, $\delta_{\mathrm{kin}}=1.8\pm0.1$, are lower than the ones obtained from the brightness temperature measurements, $\Gamma_{\mathrm{T_{\mathrm{b}}}}=4.9\pm0.5$ and $\delta_{\mathrm{T_{\mathrm{b}}}}=9.7\pm1.0$ for the core, and $\Gamma_{\mathrm{T_{\mathrm{b}}}}=1.8\pm0.2$ and $\delta_{\mathrm{T_{\mathrm{b}}}}=3.4\pm0.3$ for the hotspot. The difference between the Doppler factor obtained from kinematics and the brightness temperature suggests that the high $T_{\mathrm{b, obs}}$ in the hotspot is not the result of relativistic beaming, but is due to high intrinsic brightness temperature, likely arising from the interaction of the jet with the ambient medium \citep[see Sect.~\ref{morphology} and][]{2021ApJ...920L..24K}. The measured brightness temperatures as well as the estimated Lorentz and Doppler factors are shown in Table~\ref{tab:Tb}.

\begin{figure}[h!]
    \centering
    \includegraphics[width=0.9\linewidth]{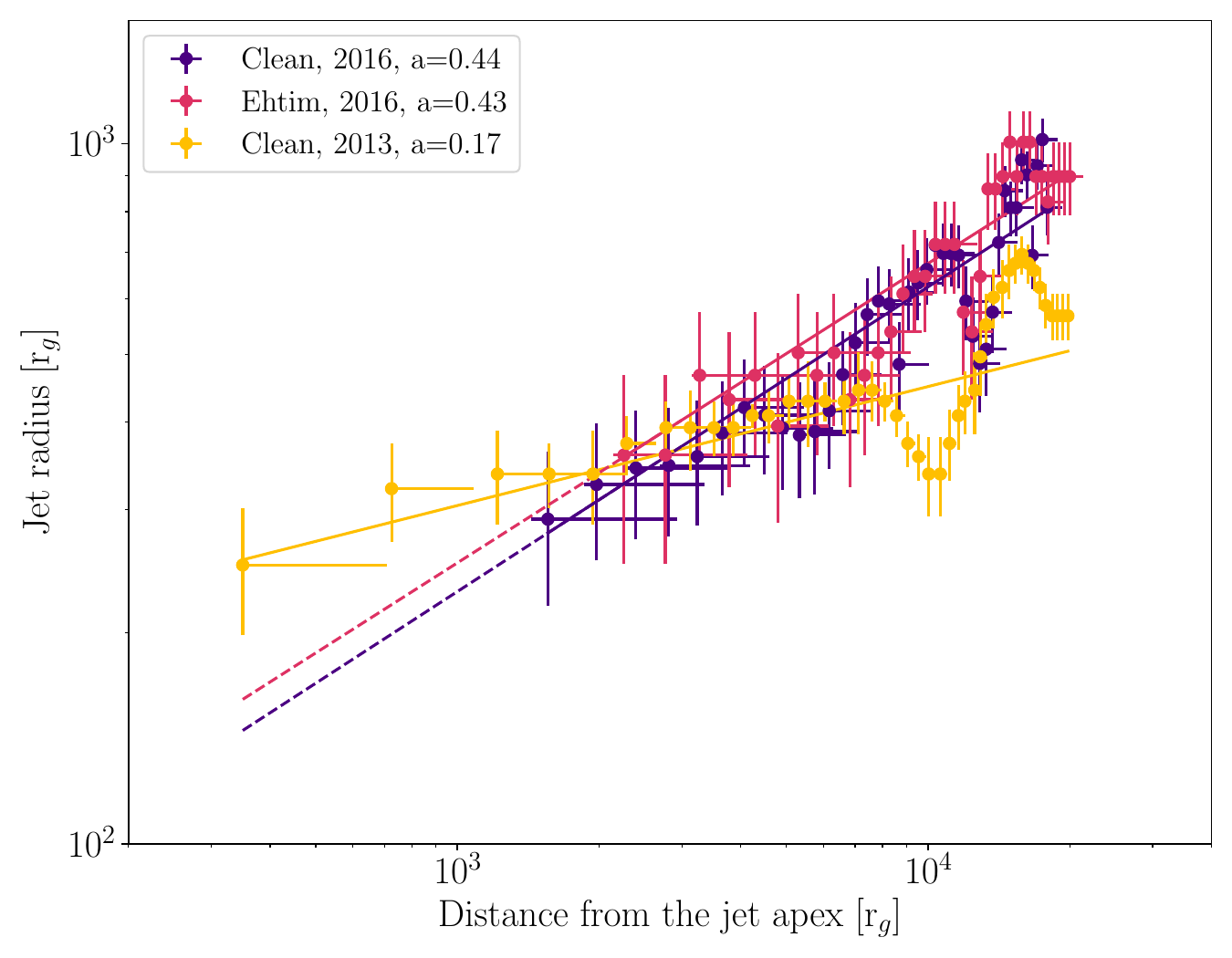}
    \caption{Collimation profile of 3C\,84 showing the jet width as a function of de-projected distance from the jet apex for the $2013$ and $2016$ measurements using the core shift constraint based on the jet--counter-jet distance. For the latter, we plot the collimation profiles obtained from both the clean image and the \texttt{ehtim} image. The jet width is measured as the separation between the edge-brightened structures of the jet. The closest measurement to the jet apex in 2016 is at $\sim1560$~r$_{\mathrm{g}}$, where the jet width is $\sim290$~r$_{\mathrm{g}}$. We fit the profiles with a power-law, and the resulting power-law index is $a=0.44\pm0.07$, indicating a parabolic profile similar to that of M\,87.}
    \label{fig:coll}
\end{figure}

\section{Evolution of the parsec-scale jet}
\label{morphology}

The 22\,GHz RadioAstron high-resolution observation presented in this work allowed us to investigate the evolution of the restarted jet that emerged in 2003 \citep{suzuki12} at very high resolution. The results were compared with previous observations (see Fig.~\ref{fig:over}) carried out with RadioAstron by \citet{giovannini18}. The most noticeable changes between the two epochs are the shift of the hotspot's position and the change in the jet base's direction. The new image confirms the existence of the limb-brightened jet, as well as the limb-brightened counter-jet that was seen on the 2013 map (see Fig.~\ref{fig:cln} and ~\ref{fig:ehtim}). This confirmation is important because although the detection of the edge-brightened jet was robust, the same was not true for the limb-brightened counter-jet. The ($u$,$v$) coverage on the space baselines is very 1D both in 2013 and 2016, which could have created image artifacts that affected the structure visible on the image. However, the orientation of the space baselines is orthogonal in the two observations, which makes the detection of the counter-jet more robust, as it is recovered in both datasets with different ($u$,$v$) coverage.

In the 2016 image (Fig.~\ref{fig:over}), we see a change in the jet direction. In this epoch, the jet is more aligned toward the hotspot at 3\,mas to the south. The base of the jet has changed direction by $32\degr\pm1\degr$ between the two observations, corresponding to a $\sim11\degr$ change per year. We measured this by quantifying the change of the spine's direction within the inner $\sim0.2$~mas of the jet. \citet{2025A&A...696A..17F} have recently shown that the direction of the jet base in the 43\,GHz VLBA maps experiences irregular oscillations of tens of degrees on a timescale of a few years. Our results are consistent with this. The irregularity and short timescale of the oscillations indicate that they probably originate in the inner accretion disk. One potential explanation could be related to Lense--Thirring torques due to a disk tilted with respect to the black hole spin \citep{2018MNRAS.474L..81L, 2023ApJ...955...72K}.

In addition, \citet{2024ApJ...970..176K} has noted that while the global propagation direction of the C3 component has been stable over two decades, it occasionally exhibited movement in the transverse direction, which can be interpreted to be due to jet precession or a jet moving through the inhomogeneous ambient medium.

The AGN jets are expected to interact with the interstellar medium and even with stars in their host galaxies \citep[e.g., in Centaurus\,A;][]{2014A&A...569A.115M}, especially in gas-rich galaxies such as 3C\,84. \citet{2021ApJ...920L..24K} recently reported a transition from an Fanaroff--Riley (FR) type II to an FR type I morphology in the restarted jet. \citet{kino18} suggests that due to an interaction with a cold, dense clump of the interstellar medium, the position of the hotspot flipped during August-September 2015, while its peak intensity increased significantly during this period. This morphological change is also seen between the two RadioAstron epochs. Based on $43$\,GHz Korean VLBI Network and VLBI Exploration of Radio Astrometry Array and VLBA light curves of the core and the hotspot presented in \citet{kino18}, in the year following the flip the flux density of the hotspot has increased to four times its original value. It has also been shown by \citet{hodgson18} that there was a $\gamma$-ray flare originating from C3 around the time of the flip. \citet{nagai17} reported an enhanced polarized emission originating from the hotspot, which is expected in regions of the compressed magnetic field \citep{1980MNRAS.193..439L}. The authors also estimated the electron number density of the clump to fall between $4\times 10^3$~cm$^{-3}$ and $2\times 10^5$~cm$^{-3}$, placing it in the narrow-line region of the AGN or a denser part of an interstellar molecular cloud. \citet{2021ApJ...920L..24K} has shown that after the flip the jet has decelerated, and during the frustration period, the position of the hotspot has moved in a counter-clockwise direction within the lobe. When it finally broke out of the interstellar medium in 2018, thus transitioning from an FR II class morphology to an FR I class, it started to accelerate to velocities nearing the speed of light. During the frustration period, the radio flux density of the hotspot was decreasing, which stopped after the break-out of the jet. \citet{2024ApJ...970..176K} also suggest that the interaction with the inhomogeneous ambient medium and the jet break-out lead to the ejection of new jet components with superluminal apparent speeds from 2017 onward. The \textit{Fermi} light curve \citep{hodgson18} shows a steady increase in photon flux after the detection of the source in 2008, which culminated around the jet break-out in a flare, after which the flux remained constant. The data presented in this paper is taken at the beginning of the jet frustration period \citep{2021ApJ...920L..24K} of $2016.7-2018.0$ when a dense clump of the ambient medium halted jet propagation. 

Brightness temperatures both in the core and hotspot significantly exceed the equipartition brightness temperature, $T_{\mathrm{eq}}\sim5\times10^{10}$~K \citep{readhead94}. However, the Doppler factor calculated from the kinematics, $\delta_{\mathrm{kin}}=1.8\pm0.1$, indicates that the hotspot emission is weakly beamed. This suggests that the hotspot, which is associated with a termination shock at the jet head, is strongly particle-dominated, probably due to the particle acceleration by the shock itself.

\section{Collimation profile measurement}
\label{collimation}

Similar to other nearby AGNs, such as M\,87 \citep{asada12}, Mrk\,$421$ \citep{giroletti06}, Mrk\,$501$ \citep{giroletti04}, Cygnus\,A \citep{boccardi16}, and Centaurus\,A \citep{janssen21}, 3C\,84 exhibits a limb-brightened structure in its jet. The structure was first observed on the 43\,GHz VLBA image by \citet{nagai14}, and it was also detected -- together with a limb-brightened counter-jet -- in both of the 22\,GHz RadioAstron observations. The collimation profile can be described as $r\propto z^a$, for which the jet width ($r$) is a function of de-projected distance ($z$) from the radio core, and the power-law index ($a$) implies a cylindrical shape for $a=0$, a parabolic shape for $a=0.5$, and a conical shape for $a=1$. Collimation profile measurements by both \citet{nagai14} and \citet{giovannini18} report a quasi-cylindrical profile with power-law indices of $0.25\pm0.03$ and $0.17\pm0.01$, respectively. While AGNs often show a parabolic profile between the radio core and the transition zone to conical expansion at $\sim10^4-10^6 r_{\mathrm{g}}$ \citep{2020MNRAS.495.3576K}, a similar, cylindrical profile with a power-law index of $0.16$ was reported in the receding jet of NGC\,1052 by \citet{2022A&A...658A.119B}. The quasi-cylindrical profile of the restarted jet can be explained by the shallow pressure profile and almost constant density of a hot cocoon surrounding the restarted jet. The low-intensity cocoon emission was indeed discovered by \citet{savolainen21} in the 5\,GHz RadioAstron data recorded simultaneously with the 22\,GHz observations in 2013. Re-collimation by a surrounding cocoon has only been studied on kiloparsec scales previously \citep{komissarov98}. The parsec-scale cocoon in 3C\,84 was likely created by the restarted jet after 2003 as shown in \citet{savolainen21}.

For our new 22\,GHz RadioAstron data we measure the jet width profile in a manner similar to that used by \citet{nagai14} and \citet{giovannini18}, by slicing the jet perpendicular to its direction and fitting 1D Gaussians on the limb-brightened edges to measure the separation of their peak positions. To account for the change in the jet direction, at each distance we rotate the slice and measure the jet width for each rotation, taking the minimum width as the slice perpendicular to the jet direction at that given distance. For the mass of the central supermassive black hole (SMBH) and the jet inclination, we adopted the same values as \citet{giovannini18}. While it is possible that the viewing angle changed between the 2013 and 2016 observations, due to, for example, jet precession \citep{2024ApJ...970..176K}, this only influences the distance of the closest jet width measurement, but not the slope of the collimation profile. We expect that, at least for the straight jet in 2016, that the viewing angle is constant along the length of the jet. For the 2013 measurement, if the viewing angles are different  but constant upstream and downstream of the bend, the shape of the collimation profile is still quasi-conical, and only the de-projected distances from the core are affected. The resulting collimation profile is parabolic with a power-law index of $a=0.44\pm 0.07$ (see Fig.~\ref{fig:coll}). The closest measurement is located at $\sim1560$~r$_{\mathrm{g}}$ from the jet apex because we could not distinguish the two jet rails closer to the core. This is caused by the lower resolution of the 2016 observation compared to the 2013 one. The jet transverse radius closest to the core is $\sim290$~r$_{\mathrm{g}}$.

\begin{figure*}[h!]
\sidecaption
\begin{subfigure}{6cm}
    \includegraphics[width=\linewidth]{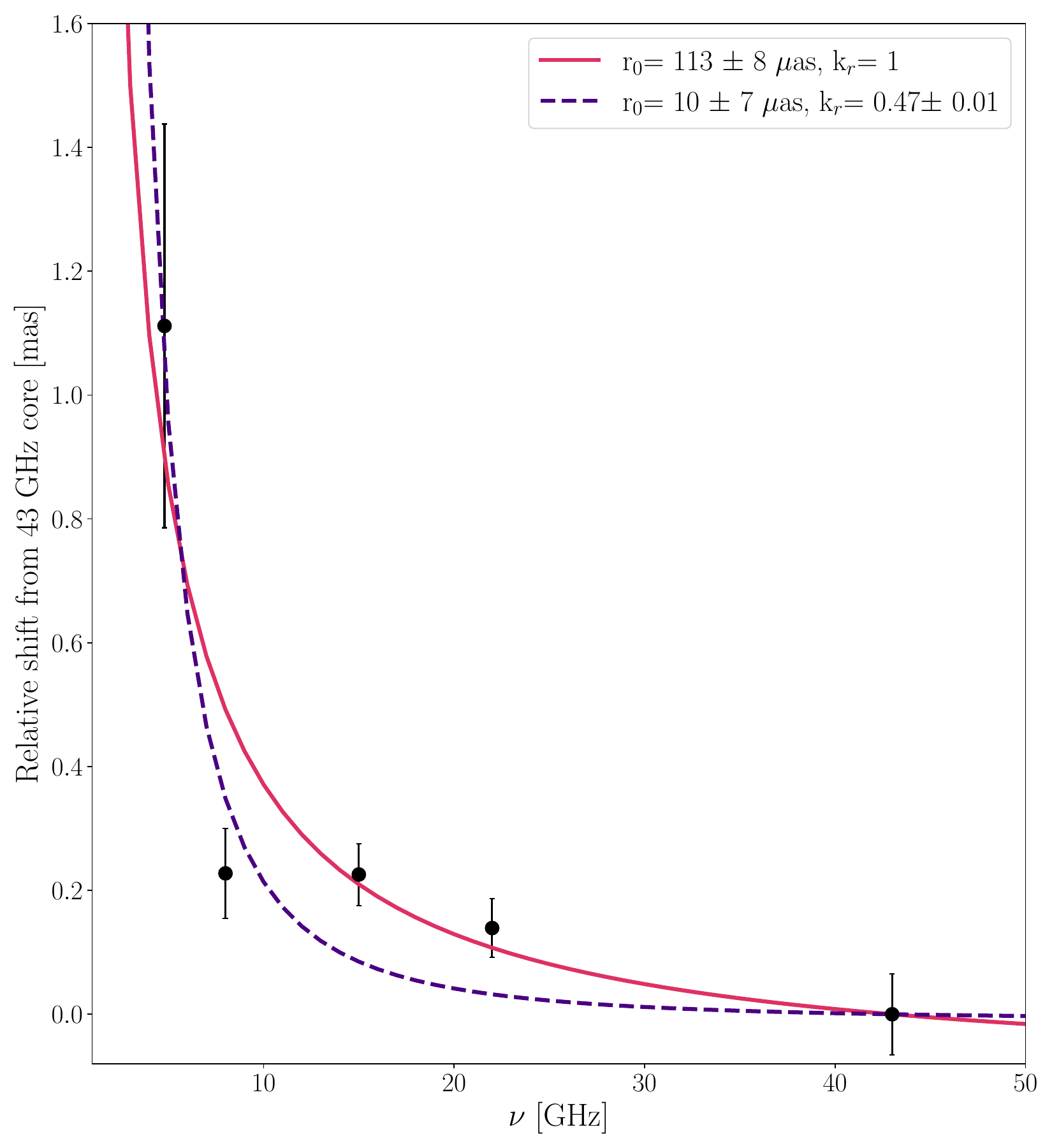}
    \end{subfigure}
    \hfill
    \begin{subfigure}{6cm}
    \includegraphics[width=\linewidth]{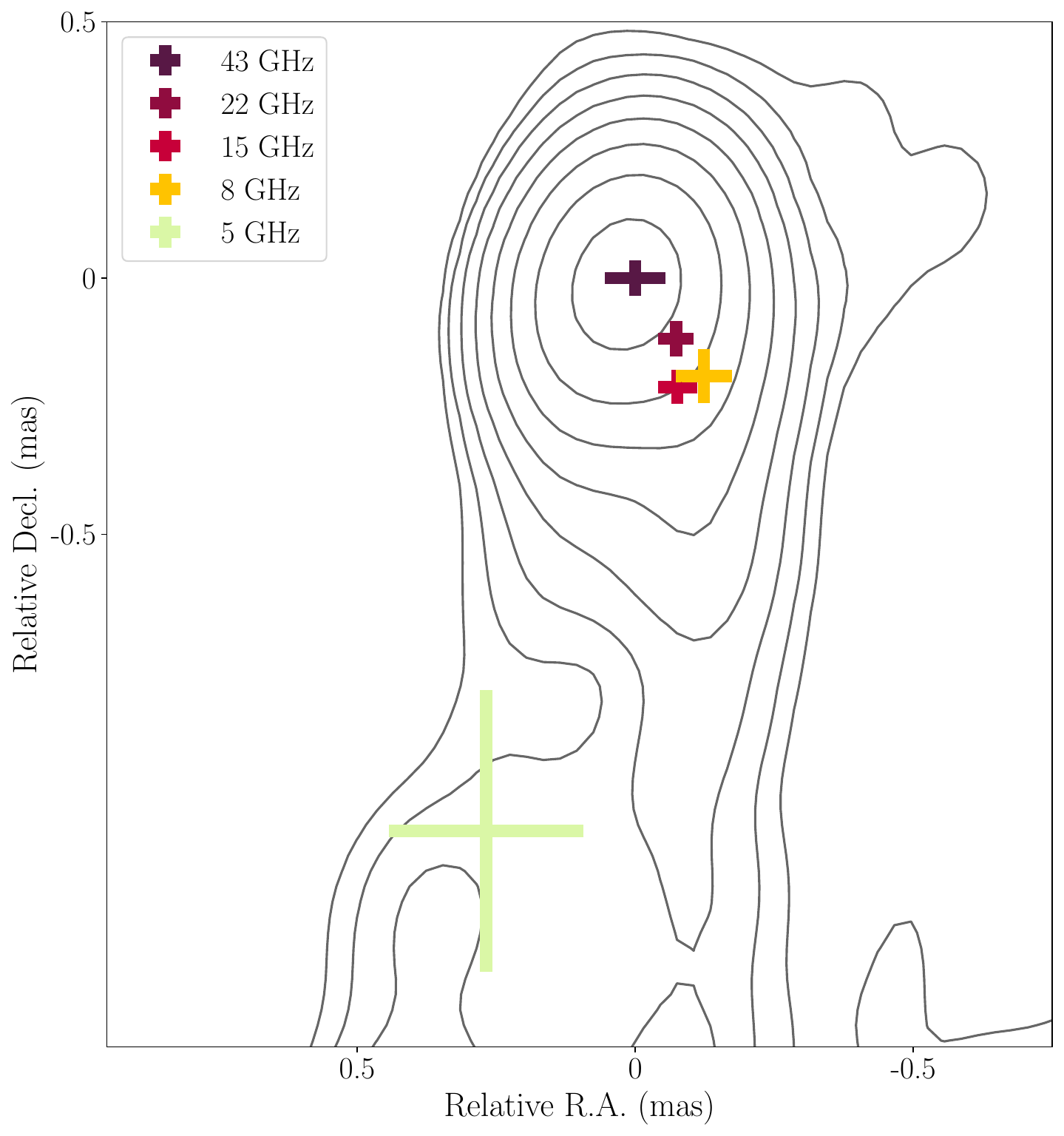}
    \end{subfigure}
    \caption{\textit{Right panel:} Core shift measurement with respect to the 43\,GHz core position. \textit{Left panel:} Positions of the lower-frequency cores with respect to the 43\,GHz one.}
    \label{fig:coreshift}
\end{figure*}

The parabolic profile, $r\propto z^{0.44\pm0.07}$ measured in this work is similar to the ones in the literature for M\,87 \citep[$r\propto z^{0.56\pm0.01}$;][]{nakamura18}, Cygnus\,A \citep[$r\propto z^{0.55\pm0.07}$;][]{boccardi16}, and many other AGNs in the MOJAVE sample \citep{2020MNRAS.495.3576K}. In a collimating, Pointing-flux-dominated jet, the collimation profile is related to the pressure profile of the external medium. If the pressure of the external medium $p_{\mathrm{ext}} \propto z^{-\kappa}$ with $\kappa < 2$, there exist an equilibrium solution with $r \propto z^{\kappa/4}$ \citep{lyubarsky09}. The change in the collimation profile between 2013 and 2016 indicates a significant steepening in the external pressure profile, from $p_{\mathrm{ext}} \propto z^{-0.68}$ to $p_{\mathrm{ext}} \propto z^{-1.76}$. Is this possible if the external pressure is provided by a hot cocoon? For a rough estimate of the cocoon pressure change, we employed the self-similar model for pressure-confined jets by \citet{kaiser97}. Assuming the jet is propagating in an ambient medium of constant density, the cocoon pressure in this model is $p_\mathrm{c} \propto Q_\mathrm{j}^{2/3} L_\mathrm{j}^{-4/3}$, where $Q_\mathrm{j}$ is the jet power and $L$ is the length of the jet. Even if $Q_\mathrm{j}$ were constant, the cocoon pressure would decrease due to increased cocoon volume as the jet head propagates. The length ratio of the 3C\,84 jet between 2013 and 2016 is 1.24, which gives a pressure ratio of 0.75. If the ambient medium density is not constant but decreases with distance, the decrease in cocoon pressure is steeper. Hence, even if the jet power does not decrease, there can be $>25$\% decrease in the cocoon pressure in 3 years, which is significant and can affect the collimation profile.

The 2013 RadioAstron observations of \citet{giovannini18} also revealed a rather wide jet base of $250$~r$_{\mathrm{g}}$ at merely $350$~$r_{\mathrm{g}}$ from the core. This led the authors to conclude that either the jet is launched from the accretion disk \citep{bp82} or the jet sheath is launched from the accretion disk and the spine from the black hole ergosphere \citep{bz77}. According to \citet{giovannini18}, it is also possible that the jet is launched from the ergosphere but expands quickly laterally in the first few hundred gravitational radii after launching. While we are only able to distinguish the limb-brightened edges at $\sim1560$~r$_{\mathrm{g}}$ from the core, we find that the jet radius is 290~r$_{\mathrm{g}}$ at this distance, which matches the 2013 jet width within the uncertainties. The change in the collimation profile shape appears to be caused by the transverse expansion of the jet at $z \gtrsim 7000$~r$_{\mathrm{g}}$. 

The 3D general relativistic magnetohydrodynamic simulations presented in \citet{2024ApJ...963L..29R} explore the collimation mechanisms that may result in a cylindrical jet width profile close to the black hole. They highlight the importance of accretion disk cooling, as such a disk collapses to the midplane as the simulation progresses and the resulting weaker disk winds allow for backflows to reach closer to the central engine than in accretion disks that do not cool efficiently. These backflows are capable of collimating the jet to a cylindrical shape closer to the black hole. However, when the disk wind starts to dominate the jet power, it will provide a stronger collimation, resulting in a smaller jet width and a quasi-parabolic collimation profile. This is another possible mechanism for collimation profile shape changes, but in this mechanism, one would expect the shape change to start from the central engine, while in 3C\,84 the transverse expansion at $z \gtrsim 7000$~r$_{\mathrm{g}}$ seems to be the driving force.

\section{Core shift measurement}
\label{coreshift}

At a given frequency, the VLBI core represents the surface where the synchrotron self-absorption opacity is $\tau=1$, whose position is therefore frequency-dependent \citep{blandford79}. Since higher frequencies probe regions closer to the central SMBH, we can extrapolate the location of the central engine based on the relative core position shifts, as a function of frequency, with respect to one of the maps if the jet stays self-similar down to the central engine \citep{marcaide84, lobanov98a}.

The location of the SMBH can be estimated by core shift measurements. The detection of the counter-jet in the 22\,GHz RadioAstron image by \citet{giovannini18} from 2013 enabled the authors to constrain the distance between the SMBH and the 22\,GHz core to be less than 30~$\mu$as. On the other hand, multifrequency VLBI observations of \citet{paraschos21} from 2015 place the jet apex farther out, $76-90$~$\mu$as, from the 86\,GHz core. Long-term Global mm-VLBI array measurements of \citet{oh21} between 2008 and 2015 are in agreement with this, putting the jet apex between 54 and 215~$\mu$as upstream of the 86\,GHz core. Placing the SMBH this far from the radio core would suggest a narrower jet base than measured in the 22\,GHz RadioAstron image. However, the recent work by \citet{2024A&A...685A.115P} measured a core shift of 14.1~$\mu$as between their 22 and 43\,GHz maps from 2022, significantly smaller than \citet{paraschos21}.

First, we measured an upper limit on the core shift based on the detection of the jet base in the limb-brightened jet and counter-jet. Under the assumption that the location of the black hole falls between the jet and the counter-jet ridges, we interpreted the branching of the eastern side of the core as the start of the jet and counter-jet. We did this based on the method described in \citet{giovannini18}, in the section titled ‘‘Possible core-shift,’’ by obtaining a slice through the western limb of the jet where the limbs of the jet and the counter-jet branch off. We fit two 1D Gaussians to this slice and measured the distance between their means, which is $205\pm9~\mu$as, where the error is the standard error of the fit. This placed an upper limit on the core shift of $103\pm5~\mu$as with respect to the 22\,GHz core. This value is larger than the 30~$\mu$as reported in \citet{giovannini18}, most likely due to the lower resolution of the 2016 observations. However, it is also possible to measure the core shift using the multifrequency ground array data based on image alignment.

We estimated the distance to the jet apex with respect to the 43\,GHz core position. For this, we followed the method described in \citet[][]{2009MNRAS.400...26O}, \citet{pushkarev12}, and \citet{2013A&A...557A.105F}, for example. First, we aligned the maps of two adjacent frequencies on the optically thin jet using 2D cross-correlation implemented in the software \texttt{FITSalign}.\footnote{FITSalign is available at: \url{https://github.com/tsavolainen/FITSalign.git}} Before applying the cross-correlation, we matched the ($u$, $v$) range of the datasets so that they would be similar, and for the pixel and restoring beam sizes for the image pairs, we chose those of the lower-frequency map. As \texttt{FITSalign} allows the image area used for the cross-correlation to be selected, we tried to align the maps based on the optically thin lobe or jet emission, excluding the optically thick C3 component. Excluding the optically thick core region was difficult at 4.8 and 8\,GHz because the core emission becomes blended with the jet emission due to decreasing resolution, so we used the faint lobes to align these maps, which increases the uncertainty of the alignment. The decreased resolution also made obtaining the position of the core difficult, as the hotspot, which is dominating the central region at these frequencies, can drag down the core component’s position toward the south. After the alignment, we used the core component positions, which we obtained from \texttt{modelfit} in \texttt{Difmap} in order to measure the distance between the core components between the different frequencies (see Table~\ref{tab:cs_comps}). Errors in the image alignment were taken as the range of shifts that lead to satisfactory spectral index maps between a given frequency pair, and \texttt{modelfit} error positions of the core were estimated based on the method described in \citet{1976ApJ...208..177L} and previously applied to VLBI data in Sect. 4.1 of \citet{chamani21}.

If we assume a Blandford-K\"onigl-type jet where both the particle density and the magnetic field decrease with the distance and the magnetic flux is conserved, then the apparent distance between the VLBI core and the jet apex is given as \citep{blandford79,lobanov98a}
\begin{equation}
\Delta r_{\mathrm{core}} [\mu\mathrm{as}] = r_0 \Bigg[\Bigg(\frac{\nu}{43 \mathrm{GHz}}\Bigg)^{-1/k_{\mathrm{r}}} -1 \Bigg], 
\end{equation}
where $r_0$ is the distance to the jet apex from the 43\,GHz core. In the case of a conical jet in equipartition, $k_{\mathrm{r}}=1$. As the jet of 3C\,84 is not conical but parabolic, we also fit $k_{\mathrm{r}}$. An additional motivation for this is the jump between the 22 and 43 GHz core distances that breaks the $\nu^{-1}$ dependence. The positions of the lower-frequency cores with respect to the 43\,GHz one, as well as the fit results, are shown in Fig.~\ref{fig:coreshift}.

With $k_{\mathrm{r}}=1$ we obtained the jet apex location to be $113\pm8~\mu$as from the $43$\,GHz core and $252\pm34~\mu$as from the $22$\,GHz core. When $k_{\mathrm{r}}$ was fit, we obtained $k_{\mathrm{r}}=0.47\pm0.01$ and a distance to the jet origin from the 43 GHz core of $10\pm7~\mu$as and $149\pm34~\mu$as from the 22\,GHz core. The $k_{\mathrm{r}} = 1$ core shift value agrees well with what is reported in \citet{paraschos21}, who obtained the core shift using the same method as described above from data recorded in 2015. However, the shift between 22\,GHz and 43\,GHz is significantly larger than what the counter-jet argument of \citet{giovannini18} allows. This discrepancy is difficult to reconcile, considering that the 2016 RadioAstron image confirms the existence of a limb-brightened counter-jet. Therefore, we cannot exclude the possibility that the core shift measurement of this highly complex source involves unknown systematic errors. It is also important to note that the significant morphological evolution of 3C\,84, especially of the core region \citep[shown by, e.g.,][]{2024ApJ...970..176K, 2025A&A...696A..17F}, makes a direct comparison between the core shift measured via the same method at different epochs un-obvious, as changes in the jet launching region can significantly affect the core shift. We also have to mention that free-free absorption that has previously been reported in the lobes \citep{walker2000} and the restarted jet \citep{nagai17}, can affect the core shift. However, since its impact is dependent on the distribution and the homogeneity of the absorbing material, it is difficult to assess this effect.

We can also estimate the magnetic field strength based on the core-shift measurement \citep{lobanov98a, hirotani05, 2015MNRAS.451..927Z, lisakov17} and compare these magnetic field strengths with the ones obtained from our spectral fit (see Sect.~\ref{spectral}). First, we calculated $\Omega^{\nu}_{\mathrm{r}}$ between $\nu_1=43$\,GHz and $\nu_2=5$\,GHz \citet{lobanov98a}:
\begin{equation}
\label{omega}
    \Omega^{\nu}_{\mathrm{r}} [\mathrm{pc GHz^{1/k_{\mathrm{r}}}}] = 4.85\times10^{-9} \frac{\Delta r_{\mathrm{core}}D_{\mathrm{L}}}{(1+z)^2} \Bigg( \frac{\nu_1^{1/k_{\mathrm{r}}} \nu_2^{1/k_{\mathrm{r}}}}{\nu_2^{1/k_{\mathrm{r}}}-\nu_1^{1/k_{\mathrm{r}}}} \Bigg).
\end{equation}
We used the following equation from \citet{2015MNRAS.451..927Z} to calculate the equipartition magnetic field strength at $1$~pc, or in this case $2.82$~mas, from the central engine:
\begin{equation}
    B^{\mathrm{eq}}_{\mathrm{1pc}} [\mathrm{G}] \approx 0.025 \Bigg[\frac{\Omega^{3k_{\mathrm{r}}}_{r\nu} (1+z)^{3}}{\delta^{2} \phi \sin^{3k_{\mathrm{r}}-1}\theta} \Bigg]^{\frac{1}{4}},
\end{equation}
where $\phi$ is the intrinsic opening angle. We calculated $\phi$ by extrapolating the jet width profile to the projected distance of 1~pc from the core at the 22\,GHz RadioAstron image, which gave 1.90~mas. The apparent opening angle is $\phi_{\mathrm{app}}=2\arctan{(r/z)}=68\pm1\degr$, and the intrinsic opening angle is $\phi=2\arctan{(\tan{(\phi_{\mathrm{app}}/2})\sin{\theta}})$, which yields $\phi=24\pm1\degr$.

We also calculated the non-equipartition magnetic field strength at $1$~pc from the central engine \citet{2015MNRAS.451..927Z}:
\begin{equation}
    B^{\mathrm{non-eq}}_{\mathrm{1pc}} [\mathrm{G}] \approx \frac{3.35\times10^{-11}  D_L \Delta r_{\mathrm{core}} \delta \tan{\phi}}{(\nu_1^{-1} - \nu_2^{-1})^5 [(1+z) \sin{\theta}]^3 F_{\nu}^2},
\end{equation} 
where $F_{\nu}$ is the flux density in the flat, $\propto\nu^0$, part of the spectrum. $F_{\nu}$ is the flux density of the inner jet that excludes the flux density contribution of the outer lobes, so we assume $F_{\nu}$ to be the sum if the core components at 22\,GHz.

At $1$~pc or $2.82$~mas from the central engine, which falls in the jet, we measured $B^{\mathrm{eq}}_{\mathrm{1pc}}=55\pm6$~mG and $B^{\mathrm{non-eq}}_{\mathrm{1pc}}=10^{-0.41\pm0.37}$~G. The $B^{\mathrm{eq}}_{\mathrm{1pc}}$ obtained for 3C\,84 is similar to the values measured in the radio galaxy sample of \citet{2014Natur.510..126Z}.

\section{Spectral analysis of ground-based VLBI data}
\label{spectral}

Using the quasi-simultaneous observations at 4.8, 8, 15 and 43\,GHz, we were able to construct spectral index maps and perform a core shift measurement. The spectral index maps were made between two adjacent frequencies using 2D cross-correlation on the optically thin jet, excluding the C3 optically thick hotspot region whenever possible, as described in Sect.~\ref{coreshift}.

\begin{figure*}[h!]
\sidecaption
  \includegraphics[width=12cm]{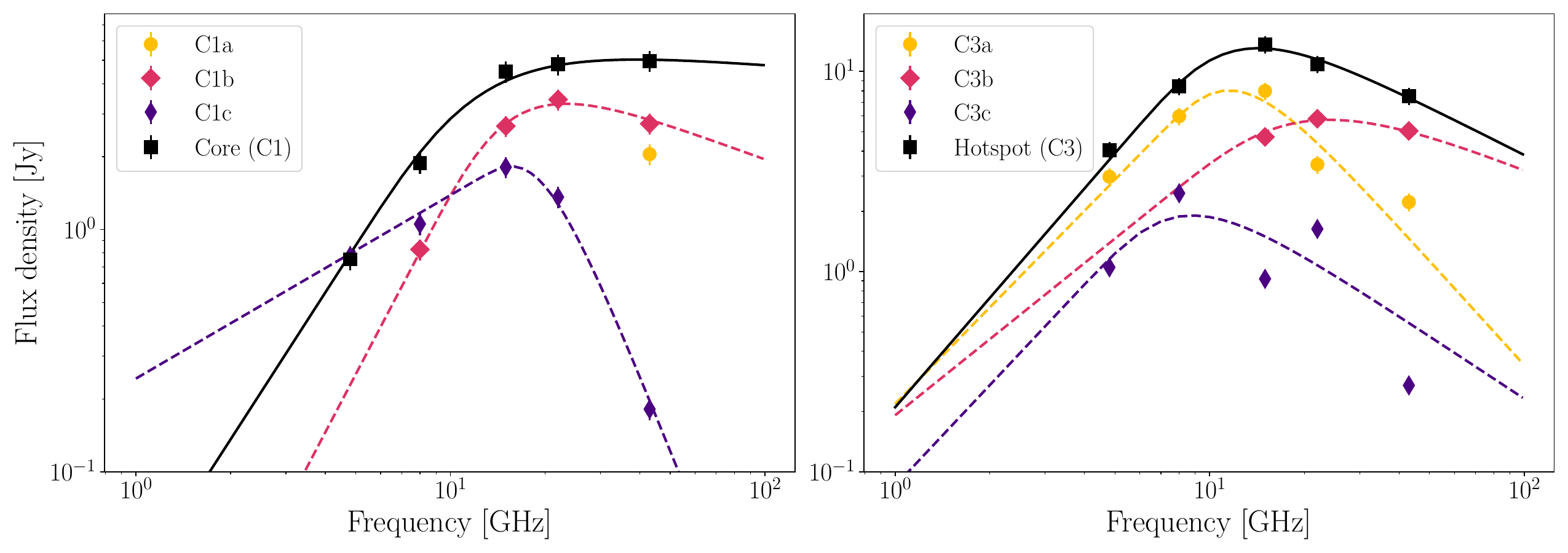}
  \caption{Spectra of 3C\,84 of the core (C1a+C1b+C1c) and jet (C3a+C3b+C3c) components and best-fit SSA models. Flux density errors are assumed to be $10$\%. Fit values are summarized in Table~\ref{tab:spect}.}
  \label{fig:spectra}
\end{figure*}

\begin{table*}[h!]
    \centering
    \caption{Self-absorbed spectrum fit parameters from Fig.~\ref{fig:spectra}.}
    \fontsize{9pt}{9pt}\selectfont
    \begin{tabular}{lccccccc}
    \hline\hline
        Component & $S_{\mathrm{m}}$ [Jy]\tablefootmark{a} & $\nu_{\mathrm{m}}$ [GHz]\tablefootmark{b} & $\alpha_{\mathrm{thick}}$\tablefootmark{c} & $\alpha$\tablefootmark{d} & $\tau_{\mathrm{m}}$\tablefootmark{e} & $a(\nu_{\mathrm{m}})$ [mas]\tablefootmark{f} & $B_{\mathrm{SSA}}$ [G]\tablefootmark{g}\\
        \hline
        C1b & $3.3\pm0.6$ & $23.3\pm4.3$ & $2.5\pm0.4$ & $-0.5\pm0.2$ & $0.3\pm0.2$ & $0.43\pm0.17$ & $10^{-0.28\pm1.09}$ \\
        C1c & $1.8\pm0.6$ & $14.4\pm3.4$ & $0.8\pm0.2$ & $-3.2\pm0.5$ & $3.8\pm0.7$ & $0.80\pm0.51$ & $10^{-0.83\pm1.68}$ \\
        core & $4.3\pm0.7$ & $16.1\pm2.8$ & $2.0\pm0.1$ & $-0.1\pm0.2$ & $0.6\pm0.2$ & $1.14\pm0.07$ & $10^{0.22\pm0.44}$ \\
        C3a & $8.0\pm0.6$ & $11.2\pm0.6$ & $1.6\pm0.3$ & $-1.7\pm0.3$ & $0.8\pm0.2$ & $0.69\pm0.36$ & $10^{-2.04\pm1.37}$ \\
        C3b & $5.7\pm0.7$ & $23.6\pm4.0$ & $1.3\pm1.0$ & $-0.6\pm0.6$ & $0.8\pm0.8$ & $0.42\pm0.25$ & $10^{-1.12\pm1.59}$ \\
        C3c & $1.9\pm0.5$ & $8.6\pm2.7$ & $1.7\pm1.0$ & $-1.0\pm0.6$ & $1.0\pm0.6$ & $0.78\pm0.41$ & $10^{-1.30\pm1.58}$ \\
        hotspot & $13.0\pm0.6$ & $14.1\pm0.8$ & $1.8\pm0.3$ & $-0.8\pm0.2$ & $0.7\pm0.2$ & $1.40\pm0.07$ & $10^{-0.39\pm0.18}$ \\
    \hline
    \end{tabular}
    \label{tab:spect}
    \tablefoot{
    \tablefoottext{a}{Flux density at the turnover.}
    \tablefoottext{b}{Turnover frequency.}
    \tablefoottext{c}{Optically thick spectral index.}
    \tablefoottext{d}{Optically thin spectral index.}
    \tablefoottext{e}{Optical depth.}
    \tablefoottext{f}{Size of the emission region at the turnover, which contains the correction factor.}
    \tablefoottext{g}{Magnetic field strength at the turnover.}
    }
\end{table*}

In the spectral index maps shown in Fig.~\ref{fig:spix}, the inverted spectrum of the counter-jet is probably due to free-free absorption, which was previously reported both in the lobes originating from an earlier outburst \citep{walker2000} and for the restarted jet as well \citep{nagai17}. The spectral index of the hotspot between the lower frequency map pairs is inverted, but between $15-22$ and $22-43$ GHz pairs, it is flat, while the jet has a slightly steeper spectral index. Spectral index maps show a north-south gradient in the core, which can be explained by synchrotron self-absorption, and an inverted spectrum in the counter-jet, which is indicative of free-free absorption. 

To ensure that we extracted the spectra from the same source regions through our observing frequencies, we utilized the model transfer method described in \citet{2006PhDT.......491S} and \citet{savolainen08}. This method allowed us to measure the flux density of components smaller than the beam size at the lower frequencies. First, we obtained a simplified model of the 43 GHz map using circular Gaussian components. To transfer these components to the lower frequency maps, we used the shifts obtained from the 2D cross-correlation to align the components between the frequencies. 

In the first round, we only fit for the flux density and size of the components. While we assume that the size of the components is either constant or shows a smooth variation with frequency, this was not the case for the 5 and 8 GHz components, which all showed a steep increase in size compared to the smooth changes between 15 and 43 GHz. As a result of this, we only fit linear trends to the 15, 22, and 43 GHz component sizes to remove the low-frequency outliers. This step was important because the component size is used for the magnetic field strength estimation, where it is assumed that the components describe the same jet region. Finally, the angular sizes were fixed to the values derived from the linear fits, and only the flux density was allowed to vary in the subsequent model fitting step. The component characteristics are listed in Table~\ref{tab:SSA_comps}.

In general, the core region could be described with two (except at 43\,GHz, where it consists of three), and the hotspot with three components, which we used to estimate the size and flux density contained within these regions. We studied the spectrum of 3C\,84 by fitting the flux density of the individual core and hotspot components, and their sum, with a synchrotron self-absorbed (SSA) spectrum \citep[][]{ssa70}:
\begin{equation}
    S_{\nu}(\nu) = S_{\mathrm{m}} \Bigg(\frac{\nu}{\nu_{\mathrm{m}}}\Bigg)^{\alpha_{\mathrm{thick}}} 
\frac{1-e^{-\tau_{\mathrm{m}} (\frac{\nu}{\nu_{\mathrm{m}}})^{\alpha-\alpha_{\mathrm{thick}}}}}{1-e^{-\tau_{\mathrm{m}}}},
\end{equation}
where $\nu_{\mathrm{m}}$ is the turnover frequency, $S_{\mathrm{m}}$ and $\tau_{\mathrm{m}}$ are the maximum flux density and the optical depth at $\nu_{\mathrm{m}}$, $\alpha$ and $\alpha_{\mathrm{thick}}$ are the spectral indices at the optically thin and thick parts of the spectrum. With the exception of the summed core component, we approximated $\tau_{\mathrm{m}}$ \citep{1999A&A...349...45T}:
\begin{equation}
    \tau_{\mathrm{m}} \approx \frac{3}{2} \Big(\sqrt{1-\frac{8\alpha}{3\alpha_{\mathrm{thick}}}}-1 \Big).
\end{equation}
Spectra are shown in Fig.~\ref{fig:spectra}, and fit parameters are summarized in Table~\ref{tab:spect}.

If the components in question are resolved, under the assumption that the emission region is homogeneous and spherical, we can estimate the synchrotron self-absorption magnetic field based on Eq.~$2$ from \citet{marscher83}:
\begin{equation}
\label{eq:ssa}
    B_{\mathrm{SSA}} [\mathrm{G}] = 10^{-5}b(\alpha)a(\nu_{\mathrm{m}})^4 \nu_{\mathrm{m}}^5 \Bigg(\frac{S_{\mathrm{m}}\tau_{\mathrm{m}}}{1-e^{-\tau_{\mathrm{m}}}}\Bigg)^{-2} \Bigg(\frac{\delta}{1+z}\Bigg),
\end{equation}
where $b(\alpha)$ can be found in \citet{marscher83} for typical spectral index values, $a(\nu_{\mathrm{m}})$ is the size of the emission region at the turnover, and can be estimated from linear interpolation based on the relation $a\propto\nu^{-1}$. In order to approximate the size of a partially opaque spherical source, the size of the fit Gaussian components was multiplied by 1.6 as the components are (partially) optically thick. The correction factor is already included in the values shown in Table~\ref{tab:spect}. The size of the core and hotspot regions across the observing frequencies were measured by adding the FWHM of their subcomponents in quadrature, and then these values were interpolated to the turnover frequency. The error of $B_{\mathrm{SSA}}$ was estimated by fitting 10000 realizations of the function, drawing from a Gaussian distribution whose mean corresponded to the value of a given parameter, with a standard deviation that corresponded to the error of said parameter. The $B_{\mathrm{SSA}}$ values we report in this paper (see Table~\ref{tab:spect}) are expectation values from the Monte Carlo simulation and differ from the values directly calculated from Eq.~\ref{eq:ssa}, as this formula is highly nonlinear.

The strength of the magnetic field in the core is on the order of 0.2$-$1.7~G (C1b and C1a+b+c combined), and for the hotspot components, the measurements fall between 9~mG (C3a) and 0.4~G (C3a+b+c combined). The combined spectrum above the turnover frequency is flat in the core, and steeper in the hotspot. A flat core spectrum indicates that the core region can be represented with a Blandford-K\"onigl type jet, while the hot spot spectrum is consistent with a more homogeneous emission region. Since C1b is the component closest to the jet apex for which it is possible to measure the turnover frequency, we compare its magnetic field strength to the $B_{\mathrm{eq}}$ obtained from the core shift. $B=B_{\mathrm{1pc}}r_{\mathrm{1pc}}/r$ is $0.62\pm0.10$~G at the distance of the 22\,GHz core, which agrees well with the $10^{-0.28\pm1.09}\approx0.5$~G measured for the C1b component.

\citet{2024A&A...682L...3P} measured a magnetic field strength of $2.9\pm1.6$~G at the turnover frequency of $113\pm4$~GHz based on the sub-milliarcsecond scale VLBI spectrum between 15 and 228\,GHz in April, 2017. This is the same order of magnitude as our value for the core region, $B_{\mathrm{SSA}}\sim1.7$~G at the turnover frequency of $16.1\pm2.8$\,GHz. Since our combined core spectrum is flat above the turnover frequency, it is not surprising that the VLBI measurements extending beyond our highest observing frequency probe a region closer to the jet apex and find higher magnetic field strengths.

\section{Summary}
\label{conclusion}

We have presented new 22\,GHz RadioAstron space VLBI observations of 3C\,84 from 2016 that follow up on the 2013 22\,GHz RadioAstron results published by \citet{giovannini18}. Our goal was to observe the evolution of the parsec-scale jet, whose latest phase of activity started when a new, bright component emerged from the core in 2003 \citep{suzuki12}. This, together with the proximity and brightness of the source, made 3C\,84 a favorable target for RadioAstron to study the formation and evolution of parsec-scale jets.

We were able to detect fringes between the ground array and RadioAstron in four scans, resulting in baselines reaching 4.32~G$\lambda$, or 4.4 Earth diameters. This corresponds to an equivalent resolution of 58~$\mu$as. Our high-resolution image displayed in Fig.~\ref{fig:cln} confirms the existence of the limb-brightened jet and counter-jet detected in 2013. We also observed that the jet has swung to the west by approximately $32\degr$ and is now more aligned with the hotspot, which is consistent with the jet direction evolution reported by \citet{2025A&A...696A..17F}. We also noted the position flip of the hotspot toward the leading edge of the jet between the two RadioAstron observations, indicative of a termination shock as the jet hits dense clumps of the ambient medium \citep{kino18}.

The width profile of the restarted jet of 3C\,84 has also evolved between the two RadioAstron observations. For the jet width, $r$, as a power-law of the distance from the core, $r\sim z^a$, the quasi-cylindrical \citep[$a=0.17\pm0.01$;][]{giovannini18} profile of 2013 changed to be parabolic ($a=0.44\pm0.07$) by 2016. The parabolic profile suggests a more rapidly decreasing pressure as a function of radius for the external medium compared to 2013, and its development can be attributed to the decreasing pressure of the hot cocoon \citep{savolainen21}, which cannot collimate the jet efficiently as the cocoon itself expands.

In addition, we have analyzed the quasi-simultaneous multifrequency dataset observed at 4.8, 8, 15, and 43\,GHz with the VLBA and Effelsberg as well as with Medicina at the lowest frequency (see hybrid maps in Fig.~\ref{fig:ground_cln}). This multifrequency dataset, together with the 22\,GHz ground array image, made it possible to perform a spectral analysis (Figs.~\ref{fig:spix} and \ref{fig:spectra}) and measure the distance to the jet apex from the 43\,GHz core. However, obtaining a reliable position of the core at 4.8 and 8\,GHz was challenging due to the difficulty of excluding the optically thick hotspot emission at these frequencies. We measure the distance from the 43\,GHz to the jet apex to be $113\pm8~\mu$as with $k_{\mathrm{r}}=1$, while we report it to be $10\pm7~\mu$as when $k_{\mathrm{r}}=0.47\pm0.01$ is fit. From the 22\,GHz core, these values correspond to $252\pm34~\mu$as and $149\pm34~\mu$as, which are significantly larger than the 30\,$\mu$as constraint placed by \citet{giovannini18} from the 2013 RadioAstron data but complementary to the values obtained by \citet{paraschos21} from 2015 observations. The discrepancy may indicate that there are unknown systematic effects in the core shift measurement of this highly complex source. In addition, comparing the core shift at different epochs is complicated by the significant morphological evolution of 3C\,84 detailed in, for example, \citet{2025A&A...696A..17F}.

We have also calculated the magnetic field strength for the core and hotspot components based on their SSA properties, and we find that the magnetic field at the turnover frequency of $\sim16$\,GHz in the core is on the order of 1.7~G, while the hotspot magnetic field strength is likely between 0 and 410~mG. The 15$-$43\,GHz core spectrum is flat, as expected, while the hotspot shows a steep optically thin spectrum above 15\,GHz. Our core region magnetic field strength is somewhat lower than the values of $B_{\mathrm{SSA}}\sim3$~G reported by \citet{2024A&A...682L...3P}, but this is understandable given the flat core spectrum and that the measurements in that work extend to higher frequencies, up to 228\,GHz, and thus probe a region closer to the jet apex.

\section{Data availability}

Maps shown in Fig.~\ref{fig:cln} and Fig.~\ref{fig:ground_cln} are available at the CDS via \url{http://cdsweb.u-strasbg.fr/cgi-bin/qcat?J/A+A/}.

\vspace{0.5cm}
\begin{acknowledgements}
The authors would like to thank the anonymous referee for their valuable comments on the manuscript. We thank G-Y. Zhao for his constructive suggestions to improve our work.
The RadioAstron project is led by the Astro Space Center of the Lebedev Physical Institute of the Russian Academy of Sciences and the Lavochkin Scientific and Production Association under a contract with the State Space Corporation ROSCOSMOS, in collaboration with partner organizations in Russia and other countries.
Based on observations with the $100$ m telescope of the MPIfR (Max-Planck-Institut für Radioastronomie) at Effelsberg.
The National Radio Astronomy Observatory is a facility of the National Science Foundation operated under cooperative agreement by Associated Universities, Inc. 
The European VLBI Network is a joint facility of independent European, African, Asian, and North American radio astronomy institutes. 
This research is based on observations correlated at the Bonn Correlator, jointly operated by the Max Planck Institute for Radio Astronomy (MPIfR), and the Federal Agency for Cartography and Geodesy (BKG).
This research was supported through a PhD grant from the International Max Planck Research School (IMPRS) for Astronomy and Astrophysics at the Universities of Bonn and Cologne.
We acknowledge the M2FINDERS project from the European Research Council (ERC) under the European Union’s Horizon 2020 research and innovation programme (grant agreement No 101018682). 
TS was supported partly by a Research Council of Finland grant 362572.
YYK was supported by the MuSES project, which has received funding from the European Union (ERC grant agreement No 101142396). Views and opinions expressed are, however, those of the author(s) only and do not necessarily reflect those of the European Union or ERCEA. Neither the European Union nor the granting authority can be held responsible for them.
This research has made use of NASA’s Astrophysics Data System Bibliographic Services.
\end{acknowledgements}

\begin{appendix}

\section{False detection probability of fringe solutions}
\label{fdr}

\begin{figure}[h!]
    \centering
    \includegraphics[width=\linewidth]{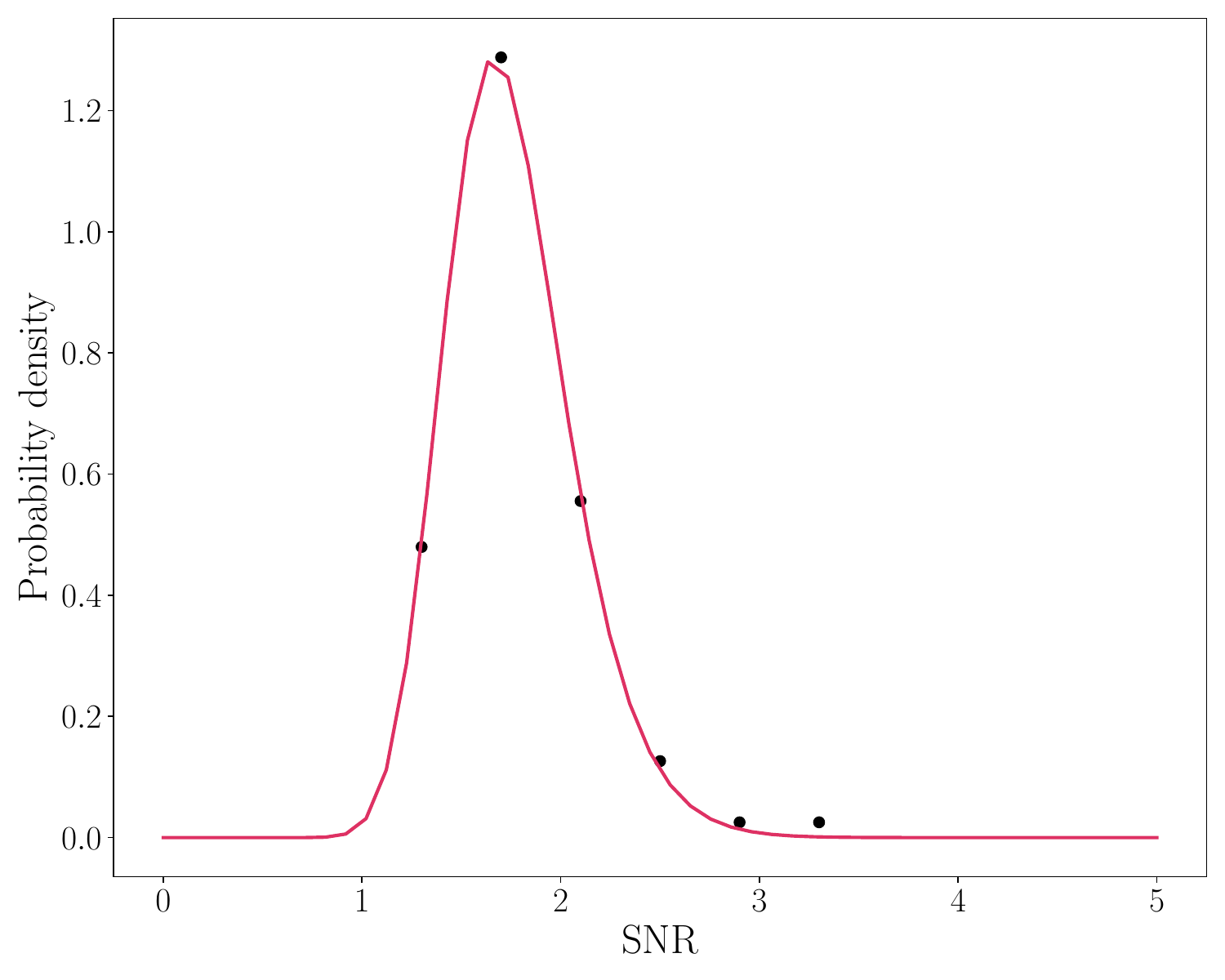}
    \caption{Distribution of the S/N of the fringe detections from one hundred fits. Search windows were $\pm100$~ns for delay and $\pm100$~mHz for rate.}
    \label{fig:fdr}
\end{figure}

Since baselines to RadioAstron have a low S/N, we calculate the false detection rates of fringe fitting on the ground--space baselines in the fast Fourier transformation (FFT) step of \texttt{FRING}. A false detection may occur when we pick up a strong noise peak as a detection or find the wrong delay and rate solution of the signal \citep{desai98} during the FFT stage of the fringe fitting. To estimate the rate of false detection, we used the following formula \citep{petrov11,savolainen21}:

\begin{equation}
    p(s)=\frac{n_{\mathrm{eff}}}{\sigma_{\mathrm{eff}}} f^2_{\mathrm{S/N}} se^{-\frac{(f_{\mathrm{S/N}}s)^2}{2}} \Bigg(1-e^{-\frac{(f_{\mathrm{S/N}}s)^2}{2}}\Bigg)^{n_{\mathrm{eff}}-1},
\end{equation}
where $n_{\mathrm{eff}}$ and $\sigma_{\mathrm{eff}}$ are the effective number of spectrum points in the search region and effective rms noise of the correlator output, and $f_{\mathrm{S/N}}$ the is a scaling factor since S/N estimates in \texttt{AIPS} are underestimated at the low S/N limit, and finally $s$ is the maximum fringe amplitude. We perform the fringe search a hundred times, adding delays and rates far outside of our search windows of $\pm100$~ns and $\pm100$~mHz for delay and rate, respectively. The probability density of the fringe S/N is shown in Fig.~\ref{fig:fdr} and false detection rates at a given S/N are summarized in Table~\ref{tab:fdr}. The S/N of the four RadioAstron fringe detections are 2.9, 3.5, 4.1, and 1.4. For the first three scans, the probability of false detection is below 1\%, while for the last scan, it is $\sim90$\%. We kept this detection because a signal was detected with \texttt{PIMA} as well, and the delay and rate changed smoothly from the previous scans (see Table~\ref{tab:delay_rate}). 

\begin{table}[h]
    \centering
    \caption{Probability of false detection at a given S/N.}
    \label{tab:fdr}
    \begin{tabular}{c c}
    \hline\hline
        S/N & P$_{e}$ \\
    \hline
         1.5 & 0.77 \\ 
         2.0 & 0.21 \\
         2.5 & 0.021 \\
         2.7 & 0.0073 \\
         3.0 & 0.0012 \\
         3.2 & 0.00033 \\
         3.5 & $4.0\times10^{-5}$ \\
         3.7 & $8.9\times10^{-6}$ \\
         4.0 & $8.0\times10^{-7}$ \\
    \hline
    \end{tabular}
\end{table}

\begin{table}[h]
    \centering
    \caption{Delay, rate, and S/N values of the RadioAstron fringe detections.}
    \label{tab:delay_rate}
    \begin{tabular}{c c c c}
    \hline\hline
        Scan~[hh:mm:ss--hh:mm:ss] & Delay~[ns] & Rate~[mHz] & S/N \\
    \hline
         23:15:00--23:30:00 & 2.48 & $-0.27$ & 2.9 \\
         23:30:30--23:45:00 & 0.00 & $-0.06$ & 3.5 \\
         23:45:30--23:59:59 & $-7.81$ & $-0.06$ & 4.1 \\
         00:00:30--00:15:00 & $-9.77$ & 0.00 & 1.4 \\
    \hline
    \end{tabular}
\end{table}

\onecolumn
\section{Multifrequency images of 3C 84}

\begin{figure*}[h!]
    \centering
    \includegraphics[width=0.85\linewidth]{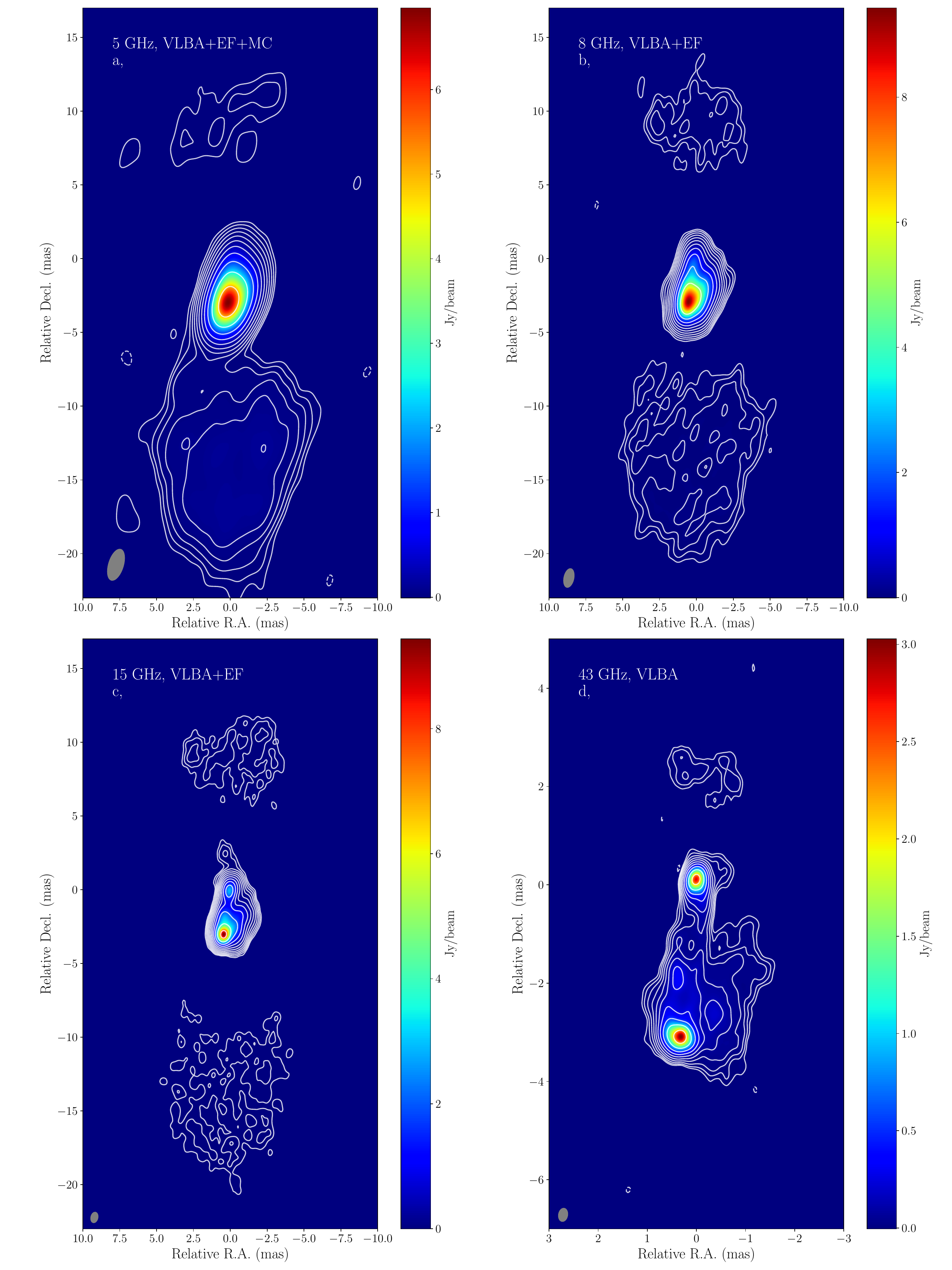}
    \caption{Clean images produced from the multi-frequency observations carried out simultaneously with the RadioAstron observations. Image properties are shown in Table~\ref{tab:ground_cln}. The maps are aligned based on the core-shift measurement, where the 22\,GHz core position is located at the map center (see Table~\ref{tab:cs_comps}).}
    \label{fig:ground_cln}
\end{figure*}

 \begin{table*}[h!]
         \centering
         \caption[]{Map properties of the clean images shown  in Fig. \ref{fig:ground_cln}.}
         \begin{tabular}{lccccccc}
            \hline\hline
            $\nu$ [GHz]\tablefootmark{a} & $S_{\mathrm{tot}}$ [Jy]\tablefootmark{b} & $S_{\mathrm{peak}}$ [Jy/beam]\tablefootmark{c} & $\sigma$ [mJy/beam]\tablefootmark{d} & $S_{\mathrm{low}}$ [mJy/beam]\tablefootmark{e} & $b_{\mathrm{maj}}$ [mas]\tablefootmark{f}& $b_{\mathrm{min}}$ [mas]\tablefootmark{g} & PA [\degree]\tablefootmark{h}\\
            \hline
            \noalign{\smallskip}
            4.8 &19.6 & 6.97 & 1.8 & 4.9 & 2.25 & 1.06 & $-17.2$ \\
            8 &  32.7 & 9.43 & 1.1 & 4.7 & 1.35 & 0.73 & $-12.4$ \\
            15 & 38.3 & 8.31 & 0.9 & 3.3 & 0.71 & 0.37 & $-16.7$ \\    
            43 & 19.5 & 3.01 & 1.0 & 4.2 & 0.29 & 0.20 & $-10.0$ \\
            \noalign{\smallskip}
            \hline
         \end{tabular}
         \label{tab:ground_cln}
         \tablefoot{
         \tablefoottext{a}{Observing frequency.}
         \tablefoottext{b}{Total flux density.}
         \tablefoottext{c}{Peak brightness.}
         \tablefoottext{d}{Rms noise.}
         \tablefoottext{e}{Lowest contour.}
         \tablefoottext{f}{Beam major axis.}
         \tablefoottext{g}{Beam minor axis.}
         \tablefoottext{h}{Beam position angle.}
         }
\end{table*}

\newpage
\section{\texttt{modelfit} components}
\label{components}

\begin{table}[h!]
    \centering
    \caption{Characteristics of \texttt{modelfit} components fit to the $22$\,GHz RadioAstron datasets.}
    \label{tab:ra_comps}
    \begin{tabular}{c c c c}
    \hline\hline
        Year &  Component & S$_{\mathrm{comp}}$ [Jy]\tablefootmark{a} & b$_{\mathrm{comp}}$ [$\mu$as]\tablefootmark{b}\\
    \hline
        2016 & core & $0.82\pm0.08$ & $32.7\pm26.2$\\
             & hotspot & $1.59\pm0.16$ & $77.1\pm26.2$\\
    \hline
    \end{tabular}
    \tablefoot{
    \tablefoottext{a}{Flux density of the component. Errors are assumed to be $10$\%.}
    \tablefoottext{b}{Size of the component. Errors are assumed to be $20$\% of the beam minor axis at the observing frequency. All components are above the resolution limit \citep{kovalev05}.}
    }
\end{table}

\begin{table}[h!]
    \centering
    \caption{Characteristics of the \texttt{modelfit} core components fit to the ground array data, used to measure the core shift.}
    \label{tab:cs_comps}
    \begin{tabular}{c c c c c}
    \hline\hline
        Frequency [GHz]\tablefootmark{a} & S$_{\mathrm{comp}}$ [Jy]\tablefootmark{b} & FWHM [mas]\tablefootmark{c} & x [mas]\tablefootmark{d} & y [mas]\tablefootmark{e}\\
    \hline
      4.8 & $0.94\pm0.09$ & $0.843\pm0.001$ & $-0.066\pm0.003$ & $2.127\pm0.004$ \\
        8 & $1.42\pm0.14$ & $0.522\pm0.001$ & $-0.457\pm0.050$ & $3.014\pm0.053$ \\
       15 & $2.67\pm0.27$ & $0.254\pm0.001$ & $-0.410\pm0.035$ & $2.993\pm0.036$ \\
       22 & $2.74\pm0.27$ & $0.209\pm0.001$ & $-0.407\pm0.022$ & $3.087\pm0.025$ \\
       43 & $3.55\pm0.36$ & $0.155\pm0.001$ & $-0.334\pm0.045$ & $3.206\pm0.025$ \\
    \hline
    \end{tabular}
    \tablefoot{
    \tablefoottext{a}{Observing frequency.}
    \tablefoottext{b}{Flux density of the component.}
    \tablefoottext{c}{Full width at half maximum of the component. All components are above the resolution limit \citep{kovalev05}.}
    \tablefoottext{d}{Position in right ascension after alignment.}
    \tablefoottext{e}{Position in declination after alignment.}
    }
\end{table}

\begin{figure}[h!]
    \centering
    \includegraphics[width=0.44\linewidth]{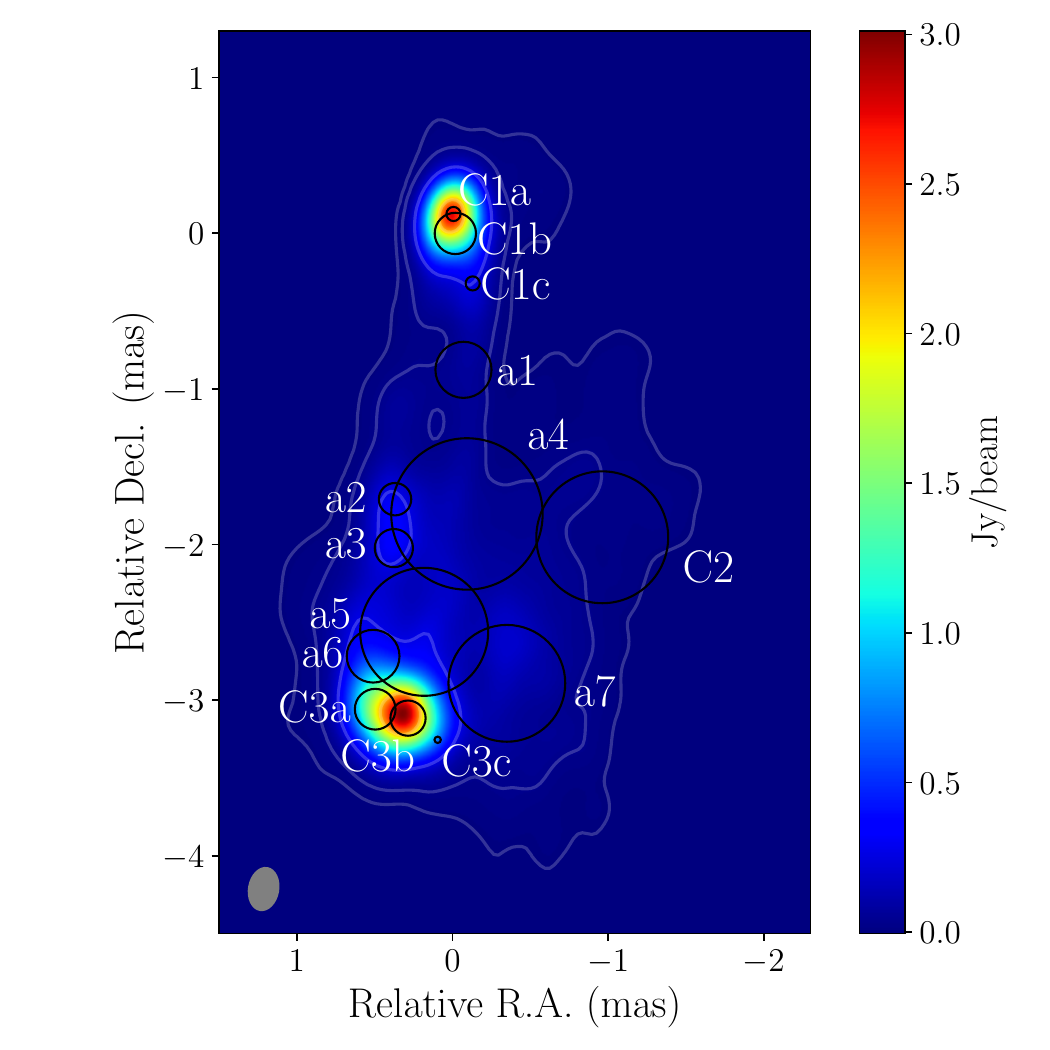}
    \caption{\texttt{modelfit} components used for the SSA fit at 43 GHz. Component characteristics are described in Table.~\ref{tab:SSA_comps}.}
    \label{fig:43mf}
\end{figure}

\begin{table}[h!]
    \centering
    \caption{Characteristics of \texttt{modelfit} components fit to the ground array data, used to perform the SSA fit.}
    \label{tab:SSA_comps}
    \fontsize{10pt}{10pt}\selectfont
    \begin{tabular}{c c c c r r}
    \hline\hline
        Frequency [GHz]\tablefootmark{a} &  Component & S$_{\mathrm{comp}}$ [Jy]\tablefootmark{b} & FWHM [mas]\tablefootmark{c} & x [mas]\tablefootmark{d} & y [mas]\tablefootmark{e}\\
    \hline
      4.8 & C1c & $0.75\pm0.08$ & $0.63\pm0.27$ & $-0.311$ & 2.466 \\
          & C3a & $1.21\pm0.12$ & $0.47\pm0.21$ & 0.316 & $-0.270$ \\
          & C3c & $0.36\pm0.04$ & $0.54\pm0.23$ & $-0.086$ &  $-0.467$ \\
          & a2 & $0.74\pm0.07$ & $0.15\pm0.09$ & 0.188 & 1.079 \\
          & a4 & $3.56\pm0.36$ & $1.54\pm0.44$ & $-0.274$ & 0.984 \\
          & a5 & $7.19\pm0.72$ & $1.48\pm0.49$ & $-0.001$ & $-0.227$ \\
          & a6 & $0.27\pm0.03$ & $0.34\pm0.07$ & 0.329 & 0.070 \\
          & a7 & $1.35\pm0.14$ & $1.13\pm0.36$ & $-0.530$ & $-0.104$\\
          & C2 & $1.08\pm0.11$ & $0.92\pm0.34$ & $-1.143$ & 0.834 \\
        8 & C1b & $0.83\pm0.08$ & $0.27\pm0.11$ & $-0.399$ & 3.087 \\
          & C1c & $1.05\pm0.11$ & $0.59\pm0.28$ & $-0.511$ & 2.766 \\
          & C3a & $5.97\pm0.60$ & $0.44\pm0.22$ & 0.116 & 0.030 \\
          & C3c & $2.46\pm0.25$ & $0.50\pm0.25$ & $-0.286$ & $-0.167$ \\
          & a1 & $0.60\pm0.06$ & $0.37\pm0.19$ & $-0.452$ & 2.210 \\
          & a2 & $1.04\pm0.10$ & $0.16\pm0.12$ & 0.012 & 1.379 \\
          & a3 & $0.58\pm0.06$ & $0.57\pm0.34$ & 0.004 & 1.065 \\
          & a4 & $4.06\pm0.41$ & $1.50\pm0.45$ & $-0.474$ & 1.284 \\
          & a5 & $9.19\pm0.92$ & $1.43\pm0.50$ & $-0.198$ & 0.527 \\
          & a7 & $3.07\pm0.31$ & $1.10\pm0.37$ & $-0.730$ & 0.196 \\
          & C2 & $1.68\pm0.17$ & $0.91\pm0.35$ & $-1.343$ & 1.134 \\
       15 & C1b & $2.67\pm0.27$ & $0.27\pm0.11$ & $-0.399$ & 3.087 \\
          & C1c & $1.81\pm0.18$ & $0.49\pm0.32$ & $-0.511$ & 2.766 \\
          & C3a & $7.99\pm0.80$ & $0.41\pm0.23$ & 0.116 & 0.030  \\
          & C3b & $4.71\pm0.47$ & $0.28\pm0.14$ & $-0.095$ & $-0.028$ \\
          & C3c & $0.92\pm0.09$ & $0.41\pm0.29$ & $-0.286$ & $-0.167$ \\
          & a1 & $0.44\pm0.04$ & $0.37\pm0.25$ & $-0.452$ & 2.210 \\
          & a2 & $0.49\pm0.05$ & $0.17\pm0.20$ & 0.012 & 1.379 \\
          & a3 & $2.30\pm0.23$ & $0.51\pm0.38$ & 0.004 & 1.065 \\
          & a4 & $5.76\pm0.58$ & $1.39\pm0.48$ & $-0.474$ & 1.284 \\
          & a5 & $4.91\pm0.49$ & $1.31\pm0.53$ & $-0.199$ & 0.527 \\
          & a6 & $0.28\pm0.03$ & $0.34\pm0.07$ & 1.291 & 0.370 \\
          & a7 & $4.39\pm0.44$ & $1.03\pm0.40$ & $-0.730$ & 0.196 \\
          & C2 & $1.03\pm0.10$ & $0.90\pm0.38$ & $-1.343$ & 1.134 \\
       22 & C1b & $3.44\pm0.34$ & $0.27\pm0.11$ & $-0.339$ & 3.087 \\
          & C1c & $1.36\pm0.14$ & $0.39\pm0.38$ & $-0.451$ & 2.766 \\
          & C3a & $3.43\pm0.34$ & $0.38\pm0.26$ & 0.176 & 0.030 \\
          & C3b & $5.81\pm0.58$ & $0.27\pm0.15$ & 0.056 & 1.065 \\
          & C3c & $1.63\pm0.16$ & $0.31\pm0.35$ & $-0.226$ & $-0.167$ \\
          & a1 & $0.28\pm0.03$ & $0.37\pm0.32$ & $-0.392$ & 2.210 \\
          & a2 & $0.25\pm0.03$ & $0.18\pm0.28$ & 0.048 & 1.379 \\
          & a3 & $1.34\pm0.13$ & $0.44\pm0.42$ & 0.056 & 1.065 \\
          & a4 & $3.36\pm0.34$ & $1.29\pm0.52$ & $-0.414$ & 1.284 \\
          & a5 & $3.17\pm0.32$ & $1.19\pm0.57$ & $-0.139$ & 0.527 \\
          & a6 & $0.56\pm0.06$ & $0.34\pm0.07$ & 0.189 & 0.370 \\
          & a7 & $2.72\pm0.27$ & $0.96\pm0.45$ & $-0.670$ & 0.196 \\
          & C2 & $0.61\pm0.06$ & $0.89\pm0.42$ & $-1.283$ & 1.134 \\
       43 & C1a & $2.05\pm0.20$ & $0.09\pm0.30$ & $-0.327$ & 3.212 \\
          & C1b & $2.74\pm0.28$ & $0.27\pm0.11$ & $-0.339$ & 3.087 \\
          & C1c & $0.18\pm0.02$ & $0.09\pm0.59$ & $-0.451$ & 2.766 \\
          & C3a & $2.23\pm0.22$ & $0.26\pm0.36$ & 0.176 & 0.030 \\
          & C3b & $5.05\pm0.51$ & $0.23\pm0.20$ & $-0.035$ & $-0.028$ \\
          & C3c & $0.27\pm0.03$ & $0.04\pm0.58$ & $-0.226$ & $-0.167$ \\
          & a1 & $0.22\pm0.02$ & $0.36\pm0.55$ & $-0.392$ & 2.210 \\
          & a2 & $0.37\pm0.04$ & $0.21\pm0.54$ & 0.048 & 1.379 \\
          & a3 & $0.55\pm0.05$ & $0.25\pm0.62$ & 0.006 & 1.065 \\
          & a4 & $1.34\pm0.13$ & $0.97\pm0.71$ & $-0.414$ & 1.284 \\
          & a5 & $1.70\pm0.17$ & $0.82\pm0.74$ & $-0.138$ & 0.527 \\
          & a6 & $0.88\pm0.09$ & $0.34\pm0.07$ & 0.189 & 0.370 \\
          & a7 & $1.35\pm0.14$ & $0.75\pm0.64$ & $-0.670$ & 0.196 \\
          & C2 & $0.24\pm0.02$ & $0.85\pm0.62$ & $-1.283$ & 1.134 \\%
    \hline
    \end{tabular}
    \tablefoot{
    \tablefoottext{a}{Observing frequency.}
    \tablefoottext{b}{Flux density of the component. Errors are assumed to be 10\%}
    \tablefoottext{c}{Full width at half maximum of the component. All components are above the resolution limit \citep{kovalev05}. Size errors are calculated from the linear fit to the 15 to 43 GHz component sizes (see Sect.~\ref{spectral}).}
    \tablefoottext{d}{Position in right ascension with the map peak brightness shifted to the map center.}
    \tablefoottext{e}{Position in declination with the map peak brightness shifted to the map center.}
    }
\end{table}

\newpage
\section{Spectral index maps}
\begin{figure}[h!]
    \centering
    \includegraphics[width=0.63\linewidth]{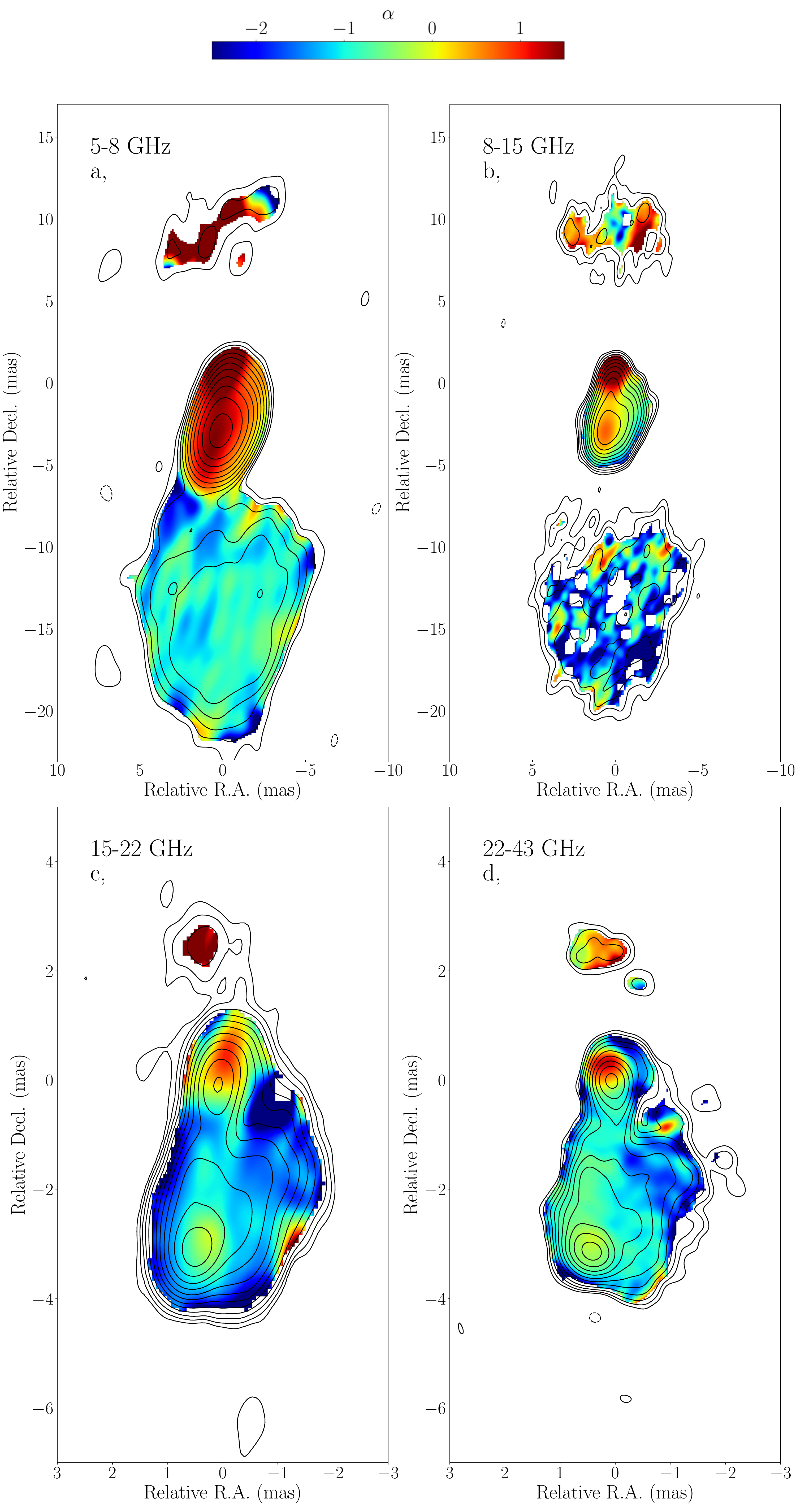}
    \caption{Spectral index maps of 3C\,84 between adjacent observing frequencies. The description of the analysis and the results can be found in Sect.~\ref{spectral}. The maps are aligned based on the core-shift measurement, with respect to the 22\,GHz core located at the map center.}
    \label{fig:spix}
\end{figure}

\end{appendix}
\end{document}